\documentclass[review]{elsarticle}

\usepackage{natbib}

\usepackage{amsmath}
\usepackage{amsfonts}
\usepackage{graphicx}
\usepackage{url} 
\usepackage{xr-hyper}
\usepackage{hyperref}
\usepackage{lscape}
\usepackage{rotating}
\usepackage{subfig}
\usepackage{multirow}
\usepackage{ltablex}
\usepackage{ragged2e}
\usepackage{booktabs}
\usepackage{verbatim}
\usepackage{placeins}
\usepackage{float}
\usepackage{graphicx}

\usepackage{booktabs}
\usepackage{longtable}
\usepackage{array}
\usepackage{diagbox}
\usepackage{wrapfig}
\usepackage{colortbl}
\usepackage{pdflscape}
\usepackage{tabu}
\usepackage{threeparttable}
\usepackage{threeparttablex}
\usepackage[normalem]{ulem}
\usepackage{makecell}
\usepackage{xcolor}
\usepackage{hhline,colortbl}

\usepackage{algorithm}
\usepackage{algorithmic}
\usepackage{afterpage}

\usepackage{bm}
\usepackage[mathscr]{eucal}

\usepackage{tikz}

\hypersetup{
    colorlinks = true,
    citecolor = {black}, 
    urlcolor = {blue}, 
    linkcolor = {black}, 
    filecolor={black}
}

\newcommand{\image}[2]{\includegraphics[#1]{#2}}

\defcitealias{KDHS2014}{KDHS, 2014}
\defcitealias{BurundiDHS:10}{BDHS, 2012}
\defcitealias{MalawiDHS:15}{MDHS, 2017}
\defcitealias{NigeriaDHS:18}{NDHS, 2019}
\defcitealias{GADM}{GADM, 2018}
\defcitealias{local2020mapping}{LBD and others, 2020}
\defcitealias{local2021mapping}{LBD and others, 2021}
\defcitealias{paige2022design}{Paige et al., 2022}
\defcitealias{dong2021space}{Dong and Wakefield, 2021}

\defcitealias{KenyaCensus:2009}{Kenya National Bureau of Statistics, 2014}

\usepackage[T1]{fontenc}
\usepackage{newpxmath}

\journal{Spatial Statistics}

\biboptions{authoryear}

\begin{document}

\begin{frontmatter}

\title{Spatial Aggregation with Respect to a Population Distribution}
%


\author[NTNUaddress]{John Paige\corref{mycorrespondingauthor}}
\cortext[mycorrespondingauthor]{Corresponding author}
\ead{john.paige@ntnu.no}

\author[NTNUaddress]{Geir-Arne Fuglstad}
\author[NTNUaddress]{Andrea Riebler}
\author[UWaddress]{Jon Wakefield}

\address[NTNUaddress]{Department of Mathematical Sciences, NTNU, Trondheim, Norway}
\address[UWaddress]{Department of Statistics and Biostatistics, University of Washington, Seattle, USA}

\begin{abstract}

Spatial aggregation with respect to a population distribution involves estimating aggregate quantities for a population based on an observation of individuals in a subpopulation. In this context, a geostatistical workflow must account for three major sources of \emph{aggregation error}: aggregation weights, fine scale variation, and finite population variation. However, common practice is to treat the unknown population distribution as a known population density and ignore empirical variability in outcomes. We improve common practice by introducing a \emph{sampling frame model} that allows aggregation models to account for the three sources of aggregation error simply and transparently.

We compare the proposed and the traditional approach using two simulation studies that mimic neonatal mortality rate (NMR) data from the 2014 Kenya Demographic and Health Survey (KDHS2014). For the traditional approach, undercoverage/overcoverage depends arbitrarily on the aggregation grid resolution, while the new approach exhibits low sensitivity. The differences between the two aggregation approaches increase as the population of an area decreases. The differences are substantial at the second administrative level and finer, but also at the first administrative level for some population quantities. We find differences between the proposed and traditional approach are consistent with those we observe in an application to NMR data from the KDHS2014.

\end{abstract}

\begin{keyword}
Geostatistics; Small area estimation; Risk; Prevalence; Bayesian; Integrated Nested Laplace Approximations.
\MSC[2020] 62H11 \sep  62P99
\end{keyword}

\end{frontmatter}


\section{Introduction} \label{sec:introduction}
Spatial aggregation based on point-referenced observations is an important problem in spatial statistics \citep{gelfand2010handbook}. 
If the quantities of interest can be written as integrals of a spatial field, the desired posterior distributions can be computed by block kriging \citep{gelfand2010handbook} or
are directly available in basis decomposition methods such as fixed rank kriging \citep{cressie:johannesson:08}, the stochastic partial differential equation (SPDE) approach \citep{Lindgren:etal:11},
and LatticeKrig and its extensions \citep{nychka:etal:15,paige2022bayesian}. However, in some cases, point-referenced measurements are collected from 
a `target' population, a finite population of interest, that may consist of people \citepalias{KDHS2014}, plants, or animal species \citep{funwi2012understanding,ballmann2017dispersal,laber2018optimal}. 
In these cases, aggregate estimates, often at multiple different areal resolutions, may be desired for the target population from which
the observations were collected. We term this problem \emph{spatial aggregation with respect to a population distribution}.

Our focus is small area estimation (SAE) of \emph{prevalence}, i.e.,~the proportion of individuals with outcome 1, based on binary responses (0 or 1). Some approaches \citep{giorgi:etal:18,dwyer2019mapping,osgood:etal:18} approximate the prevalence with the \emph{risk}, which is the expected number of individuals with outcome 1. It is worth emphasizing that even if we knew the risk in an area exactly,
say $r = 0.7$, the prevalence $p$ could vary widely around this number for a small population size just as an empirical binomial proportion might vary around its probability.

In this context, we identify three major sources of \emph{aggregation error}: 1) aggregation weights, 2) fine scale variation, and 3) finite population variation. By aggregation weights we mean the weights used to take a weighted integral or average of point level estimates to produce areal estimates. These weights may involve population density, for example, or the proportion of population in the urban or rural part of an area. Fine scale variation is variability occurring at the finest modeled spatial scale, such as the scale of the response. Fine scale variability could be induced by unmodeled nonspatial or discrete spatial covariates, for example, or other local conditions. Finite population variation is variability caused by the finite size of the target population, and is the cause of variation in prevalence about the underlying risk.

\begin{figure}
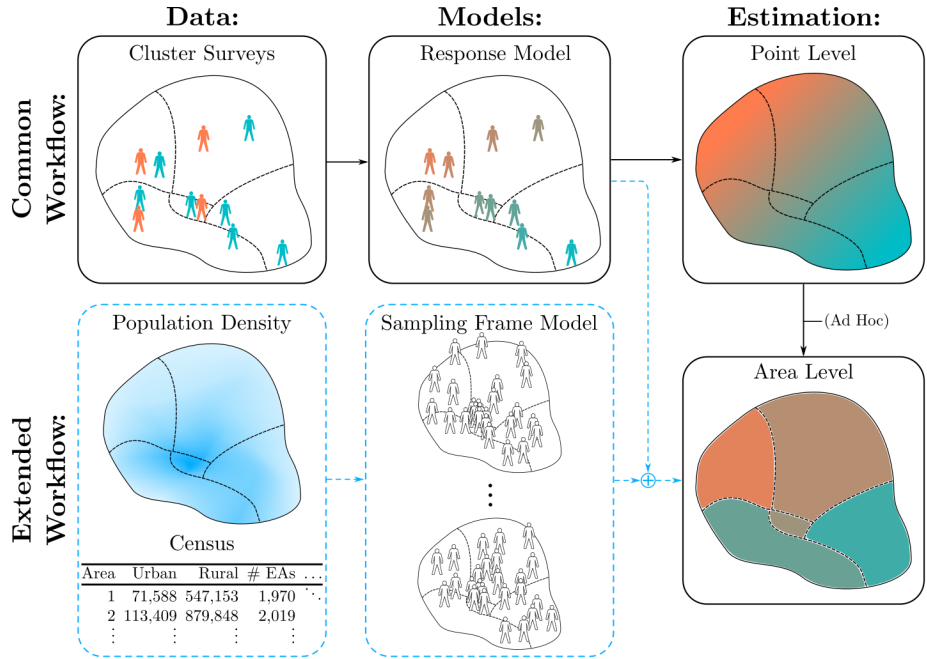

\centering
\image{width=1.0\textwidth}{illustration.png}
\caption{Common geostatistical approach to spatial aggregation with respect to a population distribution (in black), with proposed additions (in dashed blue). While population data is sometimes used aggregating from point to areal level in common approaches, this procedure is ad hoc and does not account for several sources of aggregation error.}
\label{fig:illustration}
\end{figure}

Geostatistical models applied to, for example, neonatal mortality, women's secondary education, child growth failure, and vaccination coverage, routinely do not completely account for aggregation error when aggregating from point level predictions to the areal level \citepalias{paige2022design, dong2021space, local2020mapping,local2021mapping}, as referenced in Figure \ref{fig:illustration}. For example, \citetalias{local2020mapping} and \citetalias{local2021mapping} do not include finite population variation, and include fine scale variation at the pixel level, which causes inference to depend on the grid resolution of the model. Also, \citet{paige2022design} does not account for finite population variation and similarly only partially accounts for fine scale variation. \citet{dong2021space} bases predictions on a single simulated population, which does not account for the full distribution of possible aggregation weights and target populations. We show in Section \ref{sec:nuggetEffect} that not (or even only partially) accounting for these sources of aggregation error can lead to poor predictive performance in some contexts.

We propose a spatial \emph{aggregation model} where we combine a \emph{response model} for the data with
a \emph{sampling frame model} that expresses uncertainty about the population distribution. The response model is a Bayesian hierarchical model with a data model, a process model, and a parameter model specifying priors. The sampling frame model by contrast describes the population distribution that is used when producing population aggregates. The term ``sampling frame''
is borrowed from survey statistics, and refers to the full list of the individuals and associated auxiliary information
such as spatial locations and covariate values. The 
spatial aggregation model generalizes models proposed in \citet{paige2022design} and \citet{dong2021space}, and 
is essential to account for aggregation error in the aggregate estimates. While some information about the sampling frame may be unknown, such as the exact number of unobserved 
individuals and their locations, external information such as population density maps provide a strong prior. Our proposed workflow is given in Figure \ref{fig:illustration}, which shows both the common geostaticial workflow as well as our proposed method for spatial aggregation with respect to a population.

We apply our proposed spatial aggregation model 
to neonatal mortality rates (NMRs), which is the prevalence of mortality among neonatals---children within 28 days of live birth. Data on NMR and neonatal mortality burden, the total neonatal mortality count, is observed from a Demographic and Health Surveys (DHS) survey. In DHS surveys, 
observations are made by selecting small areal units and sampling a subset of the individuals
within. These
areal units are called enumeration areas (EAs) and, e.g., the sampling frame in Kenya is divided into 96,251 EAs. We will treat the observations
as point-referenced observations of clusters. The cluster level is a natural place to include the nugget effect and the spatial 
aggregation model should be constructed based on this choice. DHS surveys are conducted under complex survey designs where inference on population averages and totals in a given set of areas is of primary interest. However, producing estimates at multiple areal resolutions is also of interest. While parts of the design will
be acknowledged in the proposed spatial model, as discussed in \citet{paige2022design}, survey statistics will not be a focus of this paper.

The observation of small areal units with the goal of making aggregate estimates for larger areas connects to the idea of basic areal units \citep{nguyen:etal:12spatial,zammit2017frk} and the modifiable areal unit problem (MAUP) 
\citep{gehlke1934certain,openshaw:1979aa}. The MAUP is still not well understood \citep{manley2014scale}, and we
are not proposing a general solution, but a practical way to address it in small area estimation based on cluster surveys, particularly in cluster surveys exhibiting fine scale variation.

Section \ref{sec:motivatingDataset} introduces the 2014 Kenya demographic health survey (KDHS2014), the dataset motivating this work. In Section \ref{sec:commonPractice} we define the statistical problem of interest along with common approaches of estimation, and discuss how they relate to the three considered major sources of aggregation error when aggregating point level predictions to the areal level for a population. We introduce a number of sampling frame models in Section \ref{sec:spatialAggregation}, and show how the robustness of different aggregation models can depend on their ability to carefully account for aggregation error in Section \ref{sec:nuggetEffect}. 
In Section \ref{sec:simulationStudy}, we conduct a simulation study to investigate
the relative importance of different sources of aggregation uncertainty at different spatial scales. Then
we investigate the practical implication for the analysis of NMR data from KDHS2014 in Section \ref{sec:application}. 
The paper ends with discussion and conclusions in Section \ref{sec:discussion}.

\section{Neonatal mortality in 2010--2014 in Kenya} \label{sec:motivatingDataset}

The 2014 KDHS provides information on a number of important health and demographic indicators including NMR. It consists of 1,582 clusters selected out of 96,215 EAs, where EAs are selected with probability proportional to the number of contained households. Within each selected EA, 25 households selected with simple random sampling form the associated cluster. The centroids of the sampled clusters are known up to a small amount of jittering, but the locations of the other EAs are unknown. Due to the stratification of the 2014 KDHS, the total number of urban and rural EAs is known within each `Admin1' area, the administrative areas immediately below the national level.

Our aim is to estimate NMR and the burden of neonatal mortality in 2010--2014 for Kenya's 47 Admin1 and 300 Admin2 areas, where Admin2 areas are the administrative areas defined just below the Admin1 level. Of the 301 Admin2 areas originally defined by Global Administrative Areas (GADM), we combine the bordering `unknown 8' and Kakeguria areas due to the small size and estimated population density of `unknown 8'. We are also interested in the relative prevalence in Admin1 and Admin2 areas, which we define as the prevalence in the urban part of the area divided by the prevalence in the rural part. 

\begin{figure}
\centering
\image{width=2.35in}{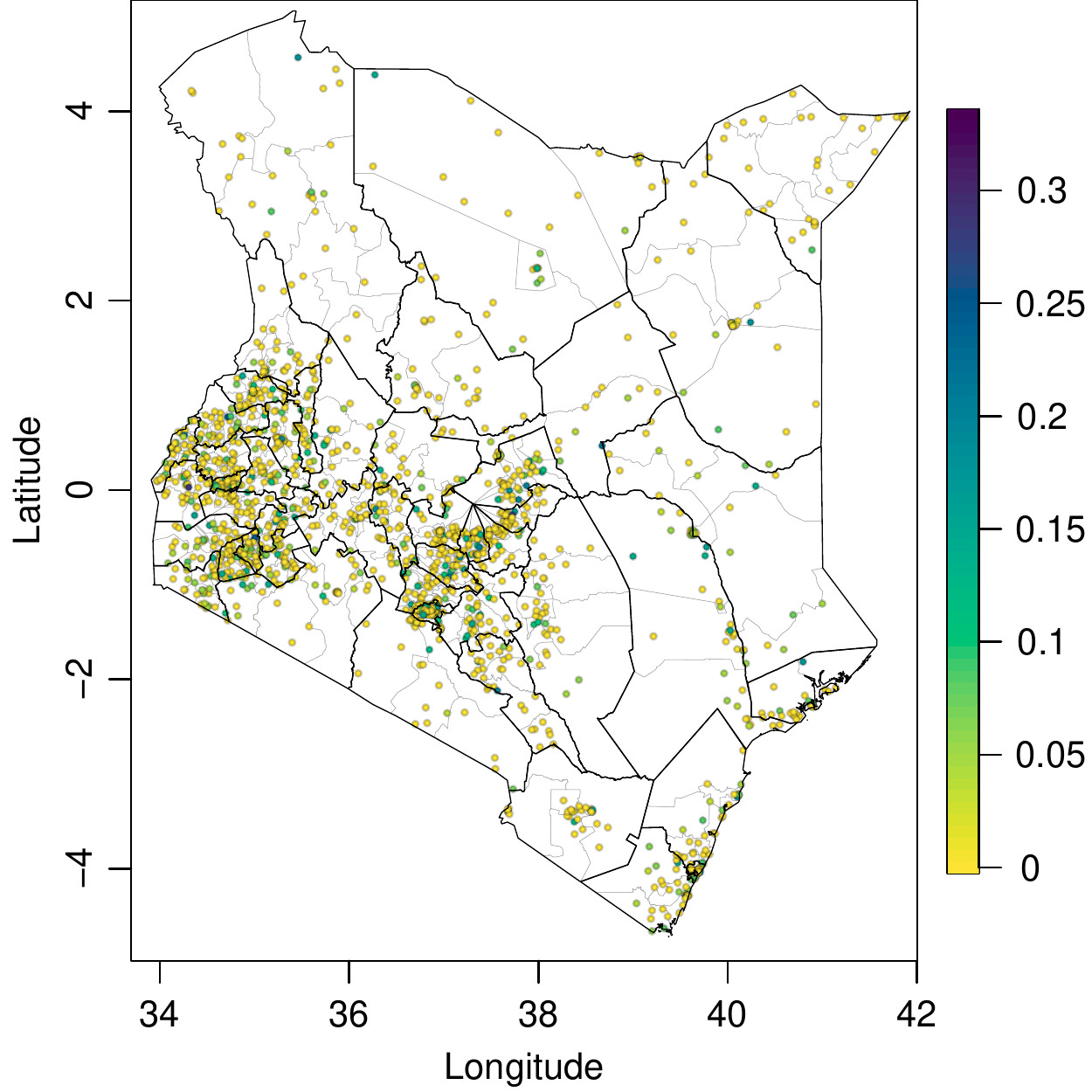} \image{width=2.35in}{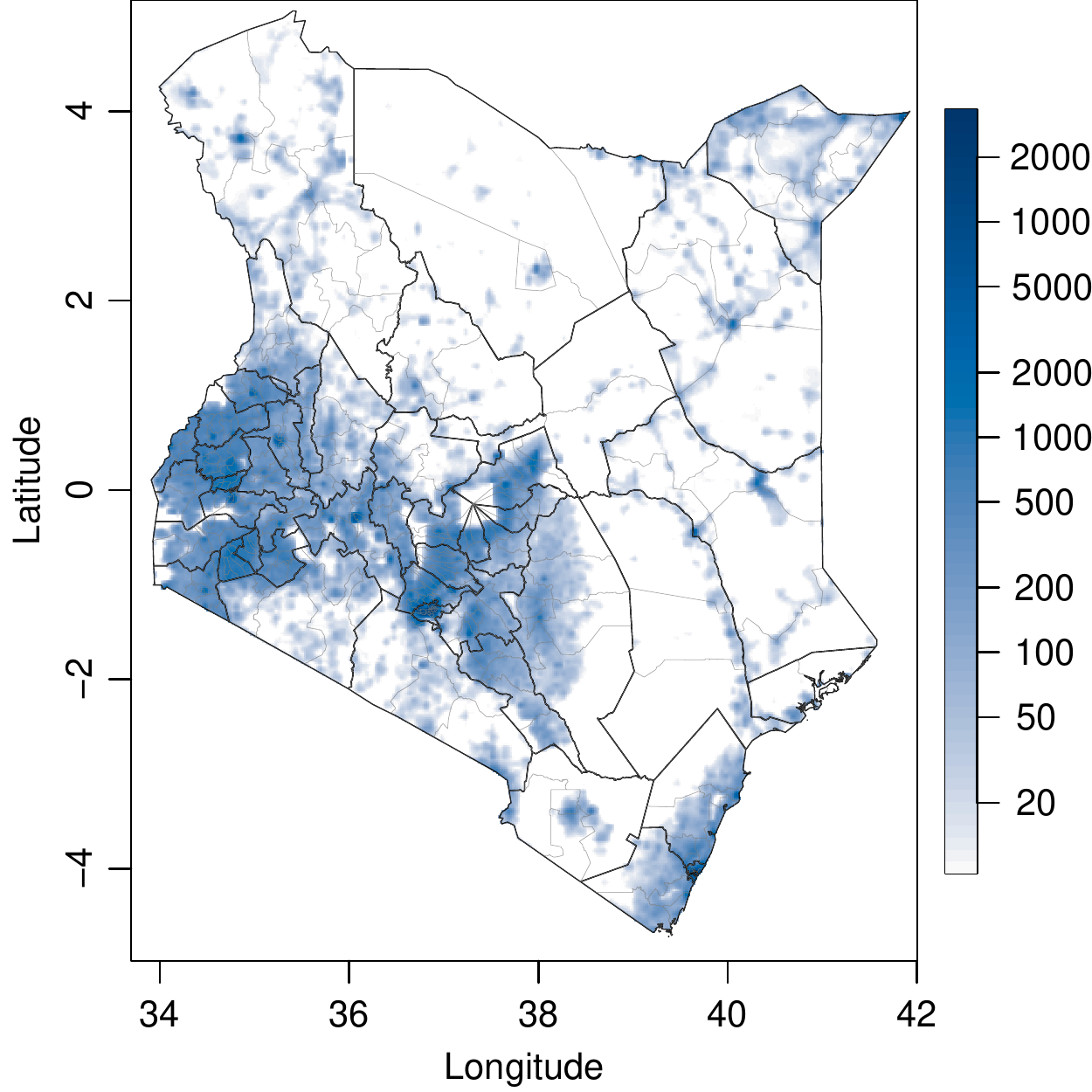}
\caption{Neonatal mortality rates in 2010-2014 as observed in 2014 KDHS sampled clusters (left), and estimated population density in people per km$^2$ in 2014 (right). Admin1 and Admin2 boundaries are shown as black and grey lines respectively.}
\label{fig:mort}
\end{figure}

Population density estimate maps are available from WorldPop \citep{stevens2015disaggregating, tatem2017worldpop} at 1km resolution, and we produce urbanicity (adjusted population density) maps by thresholding (normalizing) population density to match the urban/rural population proportions (totals) at the Admin1 level based on the 2009 census \citepalias{KenyaCensus:2009}, as in \citet{paige2022design}. More information on population density calculations and adjustments are provided in the supplement in Section \ref{sec:supplementIntegrationGrid}. Sampled NMRs are pictured for the 1,582 clusters in Figure~\ref{fig:mort} along with the estimated 2014 population density.

\section{Sources of aggregation error in common practice} \label{sec:commonPractice}

If the locations of each EA in area $A$ with index $i=1,\ldots, M$ were known along with the number of members of the target population that were born, $N_i$, and died $Z_i$ in the time period, the prevalence and burden in $A$ could be calculated as
\begin{equation}
	p(A) = \sum_{i = 1}^{M} \frac{N_{i}}{N} \frac{Z_{i}}{N_{i}}, \quad \text{and} \quad b(A) = \sum_{i = 1}^{M} Z_{i}  \label{eq:areaPrev}
\end{equation}
respectively, where $N = \sum_{i=1}^M N_i$.

Since the locations of the EAs as well as the $N_i$ and $Z_i$ are not known, a number of methods have been used to estimate prevalence that avoid the formulation in \eqref{eq:areaPrev}. Typically, these methods begin with the following standard geostatistical prevalence sampling model with notation based on \citet{diggle:giorgi:16}:
\begin{align}
	y_c \mid r_c, n_c &\sim \text{Binomial}(n_c, r_c) \nonumber \\
	\text{logit}(r_{c}) &= \boldsymbol{d}(\boldsymbol{s}_c)^T \boldsymbol{\beta} + u(\boldsymbol{s}_c) + \epsilon_c, \quad c =  1, \ldots, n. \label{eq:standardPrevalence}
\end{align}
Here, the cluster level response $y_c$ and cluster locations $\boldsymbol{s}_c$ are indexed by the cluster index $c$ for $n$ clusters in total. The cluster level risk is given by $r_c$, and the number of individuals in the target population in cluster $c$ is given by $n_c$. The vector $\boldsymbol{d}(\boldsymbol{s}_c)$ contains the spatial covariates in cluster $c$ (possibly including an intercept) with associated effect sizes given by $\boldsymbol{\beta}$, and $\epsilon_c \sim N(0, \sigma_\epsilon^2)$ is independent and identically distributed (iid) variation at the level of the response, or the spatial nugget. The spatial effect $u = \{u(\boldsymbol{s}): \boldsymbol{s} \in \mathbb{R}^2\}$ is a centered stationary Gaussian random field with marginal variance $\sigma_{\mathrm{S}}^2$.

There are various approaches for how to choose $u$, some of which are summarized in \citet{paige2022design}. For example, it is sometimes assumed that $u$ follows a reparameterization of the Besag York Molli\'{e} model known as the BYM2 model \citep{riebler:etal:16}, a solution to a stochastic partial differential equation (SPDE) \citep{Lindgren:etal:11}, or an extended LatticeKrig (ELK) model \citep{paige2022bayesian}. We will assume $u$ is stationary with an isotropic Mat\'{e}rn correlation function $\mbox{Corr}(u(\boldsymbol{s}), u(\boldsymbol{s}')) = \rho(|\boldsymbol{s} - \boldsymbol{s}'|)$ \citep{matern:86}, and will approximate it using the SPDE approach.

Since \eqref{eq:standardPrevalence} is only a model for the response, a second model is required for estimating areal prevalence. Slightly modifying notation from \citet{keller2019error}, a population quantity of interest (such as risk) for a spatial region $A$ is sometimes estimated as,
\begin{equation}
r(A) = \int_A r(\boldsymbol{s}) \ \mathrm{d} Q(\boldsymbol{s}),
\label{eq:areal}
\end{equation}
treating $r(\boldsymbol{s})$, the quantity of interest at location $\boldsymbol{s}$, as a spatial function, and where $Q(\boldsymbol{s})$ is a spatial distribution of interest. For example, the spatial density $q(\boldsymbol{s}) = \mathrm{d} Q(\boldsymbol{s})/\mathrm{d}\boldsymbol{s}$ is typically chosen as: 
\begin{itemize}
\item $q(\boldsymbol{s})=1$ for \textbf{areal totals}, 
\item $q(\boldsymbol{s})=1/|A|$ for \textbf{areal averages}, 
\item population density for \textbf{population totals}, or 
\item population density normalized to have unit integral over $A$ for \textbf{population averages} (such as for risk).
\end{itemize}
In the context of estimating population averages, sometimes areal averages are used to approximate population averages (e.g.~\citealt{diggle:giorgi:16} and \citealt{giorgi:etal:18}), and population density is based on estimates that may oversmooth in urban areas unless this effect is corrected \citep{leyk2019spatial}, such as based on census population totals. Using areal averages in lieu of population averages, and not correcting for potential oversmoothing in population density are examples of potential `aggregation weight errors', errors due to the choice in spatial distribution of interest $Q$ or in its density $q$.

Typically, \eqref{eq:areal} is approximated using a numerical integration/aggregation grid, although the resolution of the grid is not consistent. Models in this context may, for example, produce estimates at the 20km \citep{lessler2018mapping}, 5km \citep{osgood:etal:18, graetz:etal:18}, 1km \citep{utazi:etal:18}, or 100m \citep{tatem2017worldpop} resolution, and the effects of grid resolution on model estimates is not always well understood. In fact, in \citet{osgood:etal:18} and \citet{graetz:etal:18} a single nugget effect is included in each grid cell when aggregating to the areal level, weighted by the normalized population density. In that case, the number of nugget effects averaged over in an area's population average is equal to the number of grid points in the area, but the response model is the same regardless of the aggregation resolution. Hence, the variance of the population average depends arbitrarily on the aggregation grid resolution. We refer to that as the `gridded' sampling frame model, with predicted risk and expected burden in area $A$ denoted by $r_{\tiny \mbox{grid}}(A)$ and $b_{\tiny \mbox{grid}}(A)$ respectively. We call this source of aggregation error `fine scale variation' since it relates to fine scale variation in the context of the MAUP, and the nugget effect, which is sometimes interpreted as  fine scale variation.

If the ultimate goal is to estimate the population prevalence in \eqref{eq:areaPrev}, it should be emphasized that \eqref{eq:areal} is only an approximation in practice, since it typically takes the form of a continuous integral with respect to the spatial density $q$ rather than summing over the inherently discrete individuals in the population, and since $r(\boldsymbol{s})$ is the risk rather than prevalence at a point, and so does not include all the variation of prevalence. We call this source of aggregation uncertainty `finite population variation'. Despite the inherent difference between risk and prevalence, the terms are often used interchangeably \citep{giorgi:etal:18,dwyer2019mapping,osgood:etal:18}, perhaps due to an unstated assumption that finite population variation is negligible.

\section{Spatial aggregation with respect to a population distribution} \label{sec:spatialAggregation}

Our goal will be to estimate NMR in 2010--2014 using the 2014 KDHS across a set of areas, with a generic area being denoted $A$. The number of EAs in $A$ or in an area containing $A$ is known, and the number of houses and neonatals born in the time period in $A$ or in an area containing $A$ is also approximately known based on census data. We will also assume each EA has at least 25 households, since 25 households are sampled in each cluster. Although it is not too difficult to relax these assumptions, we make them for the sake of simplicity. Since the totals in the urban and rural parts of each Admin1 area are known, we can think of area $A$ as the urban or rural part of an Admin1 area, or the urban or rural parts of an Admin2 area. Given this information and the aggregation model, the joint distributions for the prevalence and burden in any set spatial region, such as for Admin2 areas or their urban or rural parts, is directly implied, even if the total target population, number of households, and number of EAs in those particular regions are all unknown.

We introduce three sampling frame models in Sections \ref{sec:empirical}-\ref{sec:smoothLatent}, for which the associated aggregation model is constructed by linking the sampling frame model to the response model. The link is formed by relating the logit risk of EA $i$ (not necessarily observed) to \eqref{eq:standardPrevalence} with $\text{logit}(r_{i}) = \text{logit}(r(\boldsymbol{s}_{i})) = \boldsymbol{d}(\boldsymbol{s}_i)^T \boldsymbol{\beta} + u(\boldsymbol{s}_i) + \epsilon_i$ (with slight modifications in the case of one sampling frame model) for each $i$, and where $\epsilon_i$ is iid $N(0, \sigma_\epsilon^2)$. Given the aggregation models share the same response model, they have the same central predictions, varying only in their uncertainty, with the empirical aggregation model having the highest predictive variance and the smooth latent model having the lowest (see Section \ref{sec:supplementRelationshipBetweenAgg} in the supplement for details). Figure \ref{fig:illustration2} provides an illustration summarizing how difference sources of uncertainty are incorporated into the main three considered aggregation models introduced in this section and their estimates of prevalence/risk.

\begin{figure}
\centering
\image{width=1.0\textwidth}{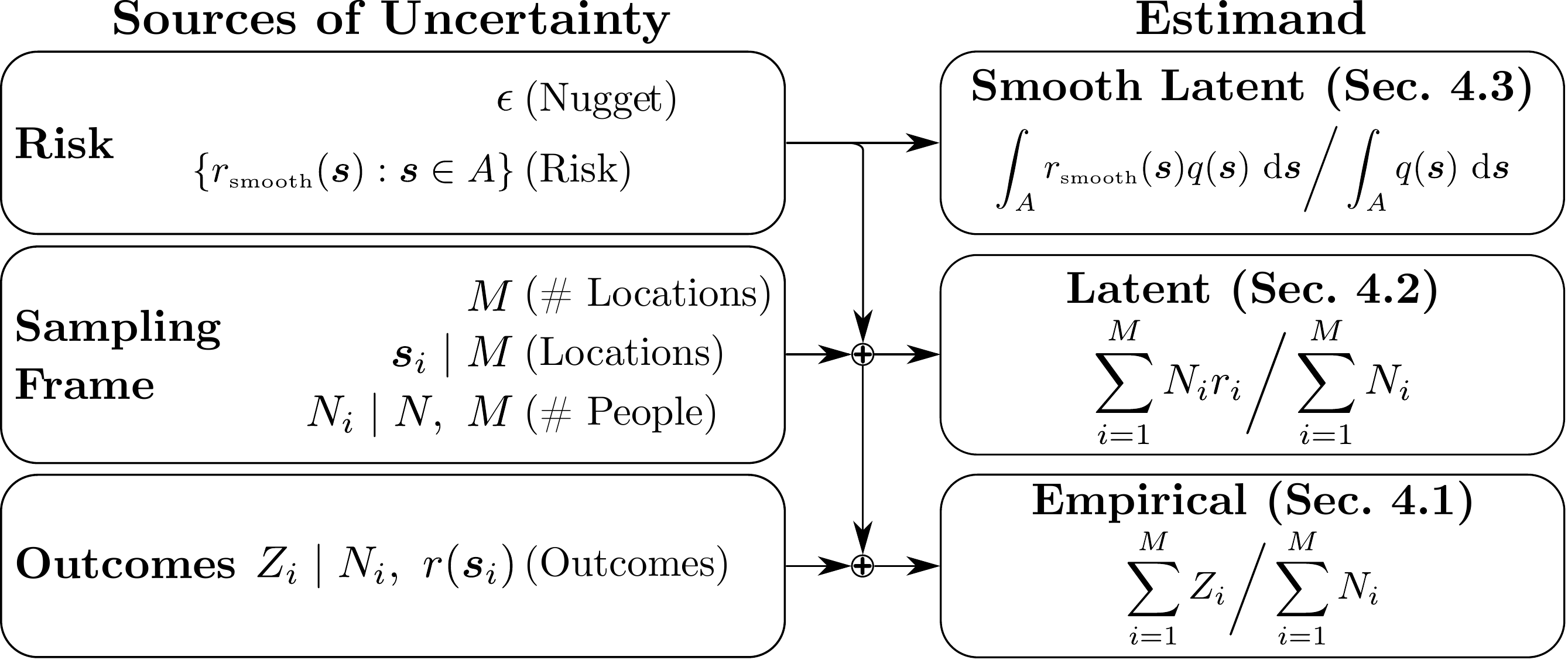}
\caption{Summary of how different sources of uncertainty are incorporated into the three main aggregation models considered: smooth latent, latent, and empirical. Model estimands for prevalence/risk along with the sections in which the models are introduced are also given.}
\label{fig:illustration2}
\end{figure}

\subsection{Empirical model} \label{sec:empirical}

Rather than approximating \eqref{eq:areaPrev} with \eqref{eq:areal}, it is possible to model the terms in \eqref{eq:areaPrev} directly. Since the number of EAs in $A$ is known, we will assume that the locations of the EAs, $\{\boldsymbol{s}_i\}_{i=1}^M$, follow a Poisson process with a possibly fixed number of points, $M$, and with rate proportional to the population density. Since $N$, the total target population and number of households in $A$, is approximately known from census data, we assume (as above) there are at least 25 households per EA, and that the rest of the total, known from census data, are distributed according to a multinomial distribution with equal probability per EA. If $h_i$ is the number of households in EA $i$ for each $i$, we distribute the total number of neonatals among EAs in $A$, $\{N_i\}_{i=1}^M$, according to a multinomial distribution with probability $h_i/\sum_i h_i$ of being in EA $i$ for each $i$. Lastly, we assume $Z_i \vert N_i, r_i \sim \mbox{Binomial}(N_i, r_i)$. We can then calculate the population prevalence and burden conditional on the $Z_i$ and $N_i$ with \eqref{eq:areaPrev}. We will denote the prevalence and burden according to the empirical model as $p_{\tiny \mbox{emp}}(A)$ and $b_{\tiny \mbox{emp}}(A)$ respectively.

We call this the `empirical' sampling frame model, since it directly models the quantities of interest, prevalence and burden, and since prevalence unlike risk is directly observable. Since population density is used in the point process model for the EA locations, and since census information is used to set the number of EAs, households, and neonatals in $A$, it correctly accounts for aggregation weights when producing the population estimate. Further, the nugget is included at the level of the EA, and $N_i$ and $Z_i$ is modeled directly for each EA, so fine scale and finite population variation are both accounted for as well.

\subsection{Latent model}

In this case, we make nearly all the same assumptions as in the empirical model, except we approximate the population prevalence with the risk, and so it is no longer necessary to assume any distribution for $Z_i \vert N_i, r_i$. To calculate the population risk and the `expected burden' with the `latent' sampling frame model, we use:
\begin{equation}
r_{\tiny \mbox{latent}}(A) = \sum_{i = 1}^{M} \frac{N_{i}}{N} r_i, \quad \text{and} \quad b_{\tiny \mbox{latent}}(A) = \sum_{i = 1}^{M} N_i r_{i}. \label{eq:areaRisk}
\end{equation}

We call this the latent sampling frame model since it only models prevalence indirectly through the risk, which is a latent quantity and cannot be directly observed. The latent model accounts for population weights and fine scale variation for the same reasons as the empirical model, but it only partially accounts for finite population variation, since it includes variation in each $N_i$ but not each $Z_i$ (except through $r_i$). This is somewhat similar to the model in \citet{dong2021space}, except it proposes a full distribution for the EA locations and population denominators rather than fixing a single draw from the distribution for the EA locations and setting $N_i \approx N/M$ for each $N_i$, rounding to the nearest integer.

\subsection{Smooth latent model} \label{sec:smoothLatent}

The `smooth latent' sampling frame model is the same as the sampling frame model used in \citet{paige2022design}. The integral in \eqref{eq:areal} is approximated numerically on a grid using spatial density $q$ equal to the normalized population density. Further, the `smooth risk' at any each grid point is calculated by integrating out the nugget (i.e.~the fine scale variation) as:
\begin{equation}
r_{\tiny \mbox{smooth}}(\boldsymbol{s}) = \int_{-\infty}^\infty \mbox{expit}\left \{ \boldsymbol{d}(\boldsymbol{s})^T \boldsymbol{\beta} + u(\boldsymbol{s}) + \epsilon \right \} \frac{\phi(\epsilon/\sigma_{\epsilon})}{\sigma_{\epsilon}} \ \mathrm{d} \epsilon,  \label{eq:smoothLatent}
\end{equation}
where $\phi$ is the standard Gaussian density. The smooth risk for an area, $r_{\tiny \mbox{smooth}}(A)$, integrates $r_{\tiny \mbox{smooth}}(\boldsymbol{s})$ with respect to population density normalized to have unit integral over $A$, and the `smooth burden', $b_{\tiny \mbox{smooth}}(A)$, can be calculated by integrating $r_{\tiny \mbox{smooth}}(\boldsymbol{s})$ with respect to the unnormalized population density.

We call this the smooth latent model because, rather than summing over discrete risks as in the latent model, the smooth risk, a continuously indexed (although possibly discontinuous) spatial function, is integrated continuously. Since the smooth latent model accounts for population density, which can be scaled to match population totals using census data as in \citet{paige2022design}, this sampling frame model does account for aggregation weight error. By integrating out the nugget effect, the smooth latent model ensures its central predictions account for fine scale variation even if it does not fully account for fine scale variation, as described in detail in Section \ref{sec:supplementRelationshipBetweenAgg} in the supplement. Since smooth risk is a risk rather than a prevalence, the smooth latent model does not account for finite population variation.

\section{Importance of aggregation assumptions: Grid resolution test} \label{sec:nuggetEffect}

We run a simple simulation study of Nairobi County NMRs based on the KDHS2014 and using a continuous spatial response model based on \citet{paige2022design} to illustrate how different areal aggregation assumptions can yield substantively different areal predictive distributions, and to show that sampling frame models that are not chosen carefully can lead to poor properties in their predictions. We do this by comparing areal predictions of four different aggregation models that share a response model, but differ in their sampling frame model and therefore in their method of producing areal estimates.

We take $\boldsymbol{d}(\boldsymbol{s}) = [1 \ I(\boldsymbol{s} \in U)]^T$ where $I(\boldsymbol{s} \in U)$ is an indicator that is 1 when $\boldsymbol{s}$ is in an urban area and 0 otherwise. We choose Nairobi County since Nairobi is the capital of Kenya, and so it is especially important to estimate its neonatal mortality prevalence accurately. In addition, it is one of the smallest and most densely populated counties in Kenya, so it is possible to increase the integration grid resolution with less computational expense.

We simulate 100 neonatal populations with associated mortality risks and prevalences down to the EA level across all of Kenya, generating one simulated survey per population, and basing simulation parameters and the survey design on the KDHS2014. This procedure is repeated for 4 grid resolutions: 200m, 1km, 5km, and 25km. In each case, we generate predictions for all 17 Admin2 areas (constituencies) in Nairobi, Kenya, calculating 95\% credible interval (CI) widths and associated empirical coverages.  For prediction, we fix the response model parameters $\boldsymbol{\theta}$ to the true values: $\sigma_{\mathrm{S}}^2=(1/3)^2$, $\sigma_\epsilon^2=1/2.5$, ${\boldsymbol{\beta} = [\beta_0 \ \beta^{\tiny \mbox{URB}}]^T= [-2.9 \ -1]^T}$, and $\rho = 400$km, where $\beta_0$ and $\beta^{\tiny \mbox{URB}}$ are the intercept and urban effect respectively. We also assume for simplicity that there are exactly 25 households and neonatals per EA. Predictions are generated for the gridded, smooth latent, latent, and empirical aggregation models, which share the same response model, and vary only in their sampling frame model.

\begin{table} \centering
\resizebox{\linewidth}{!}{
\begin{tabular}{llcrrrr}
\toprule[0.09 em]
& & & \multicolumn{4}{c}{\textbf{Grid resolution}} \\ 
& \textbf{Model} & \textbf{Units} & \textbf{200m} & \textbf{1km} & \textbf{5km} & \textbf{25km} \\ 
  \hline
95\% CI width & Empirical & (per 1,000) & 9.4 & 9.4 & 9.5 & 9.5 \\ 
& Latent && 7.8 & 7.8 & 7.8 & 7.9 \\ 
& Smooth latent && 7.2 & 7.2 & 7.3 & 7.4 \\ 
& Gridded && 8.1 & 18.8 & 56.3 & 59.7 \\ 
\addlinespace[0.3em]
95\% CI coverage & Empirical & (Percent) & 94 & 94 & 94 & 94 \\ 
& Latent && 88 & 88 & 88 & 88 \\ 
& Smooth latent && 86 & 86 & 86 & 86 \\ 
& Gridded && 90 & 100 & 100 & 100 \\ 
\bottomrule[0.09 em]
\end{tabular}}
\caption{95\% credible interval (CI) widths in neonatals per thousand, and empirical coverages in percent for considered sampling frame models as a function of aggregation grid resolution.}
\label{tab:gridResTest}
\end{table}

Table \ref{tab:gridResTest} (and Table \ref{tab:gridResTest50} in the supplement) show how 95\% (and 50\%) CI width and empirical coverage change as a function of integration grid resolution for each model. Despite the spatial response model being shared by all models, the differences in the CI widths as a function of grid resolution show how using the gridded risk sampling frame model can lead to problematic results: a 1km analysis will yield very different CI widths from a 5km analysis. In fact, at 200m resolution the gridded risk 50\% CIs achieved 46\% coverage, whereas at 25km resolution the model exhibited dramatic overcoverage with 99\% empirical coverage. This is made more problematic by the fact that there is no standard resolution at which to perform spatial aggregation. Unlike the gridded aggregation model, the empirical, latent, and smooth latent models are highly robust to the grid resolution, achieving consistent CI widths and coverages for all resolutions.

Table \ref{tab:gridResTest} and Table \ref{tab:gridResTest50} in the supplement in Section \ref{sec:supplementAdditionalGridRes} show CI widths and coverages at the 95\% and 50\% levels respectively. The empirical aggregation model consistently achieves near nominal coverage, with 94\% (52\%) coverage, whereas the latent consistently achieves 88\% (44\%), and the smooth latent achieves only 86\% (42\%) coverage for 95\% (50\%) CIs. This difference in coverage is due to the different levels of aggregation error accounted for by the models. The main difference between the latent and empirical models is that the empirical model accounts for uncertainty in the outcome, which is part of finite population variation. This suggests that finite population variation can be an important source of aggregation error.

Overall, the large differences in the empirical coverages of the aggregation models can be attributed entirely to the differences in the sampling frame models. This implies that the assumptions made when aggregating from point level predictions to the areal level matter should be considered carefully.

\pagebreak

\section{What factors influence aggregation error?} \label{sec:simulationStudy}

\subsection{Simulation setup}

To test the performance of the aggregation models under various conditions, and identify what factors influence aggregation error, we simulate from 54 different population $\times$ survey scenarios, each with a different set of simulation parameters. For each population $\times$ survey scenarios, we use the empirical aggregation model to simulate 100 populations, and then generate 100 associated surveys under a sample design intended to mimic the KDHS2014. We calculate prevalence, burden, and the relative prevalence in urban versus rural parts of areas from the simulated populations (i.e., the prevalence in the urban part of an area divided by the prevalence in the rural part). These quantities are calculated at the Admin1 and Admin2 levels as well as for the urban and rural parts of each Admin2 area when defined, and relative prevalence is only calculated at the Admin1 and Admin2 levels since it can only be calculated for areas with both urban and rural parts.

Some parameters used for the simulation study are held fixed, since they are not the focus of the simulation study. We choose a spatial range of $\rho=450$km and a nugget variance of ${\sigma_{\epsilon}^2 = 0.45}$ in order to approximately match parameters estimated for an equivalent response model as fit in Section \ref{sec:application}. We also set the urban effect to be $\beta^{\tiny \mbox{URB}} = -1$ in order to add stratification by urbanicity to ensure that predictions are able to account for stratification.

We choose $\beta_0 \in \{-4, 0\}$, and $\varphi \equiv \sigma_{\mathrm{S}}^2 / \sigma_{\epsilon}^2 \in \{1, 1/3, 1/9\}$ to be the intercept and the signal to noise ratio of the spatial response model respectively. We also choose $r_{\tiny \mbox{pop}} \in \{1/5, 1, 5\}$ to be the number of EAs (and the number of members of the population) included in the simulated population per EA in Kenya compared to the number in the KDHS2014 survey frame, and $r_{\tiny \mbox{samp}} \in \{1/3, 1, 3\}$ to be the number of clusters included in the simulated survey per cluster included in the KDHS2014. For example, if $r_{\tiny \mbox{pop}} = 1/5$, then we assume the simulated population and the simulated number of EAs are only a fifth that in the KDHS2014 sampling frame. Similarly, if $r_{\tiny \mbox{samp}} = 1/3$, then the simulated survey has only a third of the observed clusters as in the KDHS2014. The possible values of $r_{\tiny \mbox{pop}}$, $1/5$, $1$, and $5$, are chosen to roughly match the variation in the number of EA in low and middle income countries, where, for example, Nigeria has 664,999 EAs and Malawi has 12,569 EAs \citepalias{NigeriaDHS:18, MalawiDHS:15}.  The possible values of $r_{\tiny \mbox{samp}}$, $1/3$, $1$, and $3$, are chosen since DHS surveys sometimes have fewer than 400 clusters, such as in the 2010 Burundi DHS, which contains only 376 EAs \citepalias{BurundiDHS:10}, and supplementing DHS data with other sources of data is possible and is an area of active research \citep{godwin2021space, burstein:etal:18}. Overall, there are $2 \times 3 \times 3 \times 3 = 54$ different combinations of parameters, and therefore 54 different scenarios.

Throughout this work, we set a joint penalized complexity (PC) prior \citep{simpson:etal:17} on the spatial parameters $(\rho, \sigma_{\mathrm{S}}^2)$ based on \citet{fuglstad2019constructing} so that the median prior range is approximately a fifth of the spatial domain diameter, and so that $P(\sigma_{\mathrm{S}} > 1) = 0.05$. We also set a PC prior on the cluster variance satisfying $P(\sigma_{\epsilon} > 1) = 0.05$. We use INLA's default priors on the fixed effects, placing an improper $N(0, \infty)$ prior on the intercept, and an uninformative $N(0, 1000)$ prior on the urban effect.

The response models are fit to the simulated surveys, and then used in conjunction with one of three aggregation models (smooth latent, latent, and empirical) to produce estimates of prevalence, burden, and relative prevalence. Posteriors of the quantities are produced at Admin1, Admin2, and Admin2 $\times$ urban/rural levels when defined, and compared with the true population using a number of scoring rules and metrics.

\subsection{Areal prediction performance measures}

We evaluate the aggregation model performance via continuous ranked probability score (CRPS) \citep{gneiting:raftery:07}, 95\% interval score \citep{gneiting:raftery:07}, 95\% fuzzy empirical CI coverage \citep{paige2022bayesian}, 95\% fuzzy CI width \citep{paige2022bayesian}, and total computation time in minutes including fitting the response model to a single survey as well as generating all associated predictions using the aggregation model. CRPS is a strictly proper scoring rule, whereas interval scores are proper scoring rules \citep{gneiting:raftery:07}. For (strictly) proper scoring rules, the expected score of incorrect models are (strictly) larger than the expected score of the correct one, and lower scores are better. `Fuzzy' CI's \citep{geyer2005fuzzy}, are used to make empirical coverage more precise under discrete outcomes. When predicting prevalence, scores are produced on a probability scale rather than being for unnormalized population counts. Given $n_a$ areas at a given areal level, we calculate a given score as the average of the individual area scores,
\begin{align}
\mbox{CRPS}(\boldsymbol{y}^*, F_1(y^*), \ldots, F_{n_a}(y^*)) &= \frac{1}{n_a} \sum_{i=1}^{n_a} \int_{-\infty}^\infty \Big (F(y) - I(y \geq y_i^*) \Big )^2 \ \mathrm{d}y \nonumber \\
\mbox{Int}_{\alpha}(\boldsymbol{y}^*, l_1, u_1,\ldots, l_{n_a}, u_{n_a}) &= \frac{1}{n_a} \sum_{i=1}^{n_a} \bigg[ (u_i - l_i) + \frac{2}{\alpha} (l_i - y^*_i) I(y^*_i < l_i) \nonumber \\
&\quad + \frac{2}{\alpha} (y^*_i - u_i) I(y^*_i > u_i) \bigg], \label{eq:scores}
\end{align}
where $\boldsymbol{y}^* = (y^*_1, \ldots, y^*_{n_a})$ is a vector of quantities we wish to estimate (a possibly normalized version of the response), $ \hat{\boldsymbol{y}}^* = (\hat{y}^*_1, \ldots, \hat{y}^*_{n_a})$ is the vector of associated model estimates, $F_1(y^*), \ldots, F_{n_a}(y^*)$ are the associated predictive distribution cumulative distribution functions, and $l_1, \ldots, l_{n_a}$ and $u_1, \ldots, u_{n_a}$ are the lower and upper ends respectively of the predictive distribution $\alpha$-significance level CIs. We will take $\boldsymbol{y}^*$ to be the areal prevalences, burdens, or relative prevalences at the given areal level.

We also calculate relative scores. Given a proposed model, a reference model, and their respective scores $S_P$ and $S_R$, we calculate the relative score $S_{\mbox{rel}} = 100\% \times (S_P - S_R) / S_R$. In particular, we will calculate the CRPS and 95\% interval scores of the proposed empirical aggregation model relative to the smooth latent aggregation model when predicting Admin2 prevalence. We also provide other relative scores in the supplement in Section \ref{sec:supplementAdditionalSimStudy}.

\subsection{Simulation study results}
\FloatBarrier

\makeatletter
\def\clineThicknessColor#1#2#3{\@ClineThicknessColor#1\@nil{#2}{#3}}
\def\@ClineThicknessColor#1-#2\@nil#3#4{%
    \omit
    \@multicnt#1%
    \advance\@multispan\m@ne
    \ifnum\@multicnt=\@ne\@firstofone{&\omit}\fi
    \@multicnt#2%
    \advance\@multicnt-#1%
    \advance\@multispan\@ne
    {\color{#4}%
    \leaders\hrule\@height#3\hfill}%
    \cr}
\makeatother

\begin{table}[h]
\centering
\resizebox{\linewidth}{!}{
\begin{tabular}[t]{c@{\phantom{\tiny a}}c|ccc@{\phantom{a}}p{0em}ccc@{\phantom{a}}p{0em}ccc}
  & $r_{\tiny \mbox{pop}}$ & \multicolumn{3}{c}{$1/5$} &   & \multicolumn{3}{c}{$1$} &   & \multicolumn{3}{c}{$5$} \\
$r_{\tiny \mbox{samp}}$ & \diagbox[width=.3in, height=.3in, innerleftsep=.04in, innerrightsep=-.02in]{$\beta_0$}{$\varphi$} & $1/9$ & $1/3$ & $1$ &   & $1/9$ & $1/3$ & $1$ &   & $1/9$ & $1/3$ & $1$ \\
\hline
\multirow{2}{*}{$3$} & 0 & \cellcolor[HTML]{76D25A}{\textcolor{white}{\textbf{-14.6}}} & \cellcolor[HTML]{00BB81}{\textcolor{white}{\textbf{-12.1}}} & \cellcolor[HTML]{009B95}{\textcolor{white}{\textbf{-9.4}}} &  & \cellcolor[HTML]{008297}{\textcolor{white}{\textbf{-7.6}}} & \cellcolor[HTML]{006892}{\textcolor{white}{\textbf{-5.9}}} & \cellcolor[HTML]{194F86}{\textcolor{white}{\textbf{-4.3}}} &  & \cellcolor[HTML]{3F3375}{\textcolor{white}{\textbf{-2.7}}} & \cellcolor[HTML]{46236A}{\textcolor{white}{\textbf{-1.9}}} & \cellcolor[HTML]{491763}{\textcolor{white}{\textbf{-1.4}}}\\
 & -4 & \cellcolor[HTML]{D1E02E}{\textcolor{white}{\textbf{-17.1}}} & \cellcolor[HTML]{70D15C}{\textcolor{white}{\textbf{-14.5}}} & \cellcolor[HTML]{00B587}{\textcolor{white}{\textbf{-11.5}}} &  & \cellcolor[HTML]{009097}{\textcolor{white}{\textbf{-8.5}}} & \cellcolor[HTML]{007495}{\textcolor{white}{\textbf{-6.7}}} & \cellcolor[HTML]{005C8D}{\textcolor{white}{\textbf{-5.2}}} &  & \cellcolor[HTML]{3C3677}{\textcolor{white}{\textbf{-2.9}}} & \cellcolor[HTML]{45256B}{\textcolor{white}{\textbf{-2.0}}} & \cellcolor[HTML]{491763}{\textcolor{white}{\textbf{-1.4}}}\\
\addlinespace
\multirow{2}{*}{$1$} & 0 & \cellcolor[HTML]{00B586}{\textcolor{white}{\textbf{-11.6}}} & \cellcolor[HTML]{009796}{\textcolor{white}{\textbf{-9.0}}} & \cellcolor[HTML]{007696}{\textcolor{white}{\textbf{-6.8}}} &  & \cellcolor[HTML]{005F8E}{\textcolor{white}{\textbf{-5.3}}} & \cellcolor[HTML]{2E4480}{\textcolor{white}{\textbf{-3.7}}} & \cellcolor[HTML]{412E71}{\textcolor{white}{\textbf{-2.4}}} &  & \cellcolor[HTML]{481B65}{\textcolor{white}{\textbf{-1.5}}} & \cellcolor[HTML]{4A065C}{\textcolor{white}{\textbf{-0.8}}} & \cellcolor[HTML]{4B0058}{\textcolor{white}{\textbf{-0.6}}}\\[-0.33mm]  
\clineThicknessColor{8-8}{1pt}{red}
 & -4 & \cellcolor[HTML]{00C07A}{\textcolor{white}{\textbf{-12.6}}} & \cellcolor[HTML]{00B18A}{\textcolor{white}{\textbf{-11.1}}} & \cellcolor[HTML]{009696}{\textcolor{white}{\textbf{-9.0}}} &  & \cellcolor[HTML]{006691}{\textcolor{white}{\textbf{-5.7}}} & \multicolumn{1}{!{\color{red}\vrule width 1pt}c!{\color{red}\vrule width 1pt}}{\cellcolor[HTML]{115187}{\textcolor{white}{\textbf{-4.4}}}} & \cellcolor[HTML]{33417E}{\textcolor{white}{\textbf{-3.5}}} &  & \cellcolor[HTML]{481B65}{\textcolor{white}{\textbf{-1.5}}} & \cellcolor[HTML]{491662}{\textcolor{white}{\textbf{-1.3}}} & \cellcolor[HTML]{4B045B}{\textcolor{white}{\textbf{-0.8}}}\\[-0.33mm]  
\clineThicknessColor{8-8}{1pt}{red}
\addlinespace
\multirow{2}{*}{$1/3$} & 0 & \cellcolor[HTML]{008398}{\textcolor{white}{\textbf{-7.6}}} & \cellcolor[HTML]{006691}{\textcolor{white}{\textbf{-5.7}}} & \cellcolor[HTML]{055489}{\textcolor{white}{\textbf{-4.6}}} &  & \cellcolor[HTML]{3E3475}{\textcolor{white}{\textbf{-2.7}}} & \cellcolor[HTML]{44276D}{\textcolor{white}{\textbf{-2.1}}} & \cellcolor[HTML]{491763}{\textcolor{white}{\textbf{-1.4}}} &  & \cellcolor[HTML]{4B045B}{\textcolor{white}{\textbf{-0.8}}} & \cellcolor[HTML]{4B0057}{\textcolor{white}{\textbf{-0.4}}} & \cellcolor[HTML]{4B0055}{\textcolor{white}{\textbf{-0.3}}}\\
 & -4 & \cellcolor[HTML]{009597}{\textcolor{white}{\textbf{-8.9}}} & \cellcolor[HTML]{008498}{\textcolor{white}{\textbf{-7.7}}} & \cellcolor[HTML]{006791}{\textcolor{white}{\textbf{-5.8}}} &  & \cellcolor[HTML]{3B3778}{\textcolor{white}{\textbf{-3.0}}} & \cellcolor[HTML]{412F72}{\textcolor{white}{\textbf{-2.5}}} & \cellcolor[HTML]{46236A}{\textcolor{white}{\textbf{-1.9}}} &  & \cellcolor[HTML]{4B0058}{\textcolor{white}{\textbf{-0.6}}} & \cellcolor[HTML]{4B0057}{\textcolor{white}{\textbf{-0.5}}} & \cellcolor[HTML]{4B0055}{\textcolor{white}{\textbf{-0.3}}}\\
\end{tabular}}
\caption{\label{tab:PvSR_CRPS_Admin2timesstratum_prevalence}Mean percent change in CRPS of the empirical aggregation model relative to the smooth latent aggregation model, where the response is Admin2 $\times$ stratum prevalence. Yellow-green values are better, while indigo values are worse. Results most representative of the application in Section \ref{sec:application} are outlined in red.}
\end{table}

\begin{table}[h]
\centering
\resizebox{\linewidth}{!}{
\begin{tabular}[t]{c@{\phantom{\tiny a}}c|ccc@{\phantom{a}}p{0em}ccc@{\phantom{a}}p{0em}ccc}
  & $r_{\tiny \mbox{pop}}$ & \multicolumn{3}{c}{$1/5$} &   & \multicolumn{3}{c}{$1$} &   & \multicolumn{3}{c}{$5$} \\
$r_{\tiny \mbox{samp}}$ & \diagbox[width=.3in, height=.3in, innerleftsep=.04in, innerrightsep=-.02in]{$\beta_0$}{$\varphi$} & $1/9$ & $1/3$ & $1$ &   & $1/9$ & $1/3$ & $1$ &   & $1/9$ & $1/3$ & $1$ \\
\hline
\multirow{2}{*}{$3$} & 0 & \cellcolor[HTML]{D1E02E}{\textcolor{white}{\textbf{-68.5}}} & \cellcolor[HTML]{9ED947}{\textcolor{white}{\textbf{-62.9}}} & \cellcolor[HTML]{50CC68}{\textcolor{white}{\textbf{-55.8}}} &  & \cellcolor[HTML]{00BE7E}{\textcolor{white}{\textbf{-49.9}}} & \cellcolor[HTML]{00AA8F}{\textcolor{white}{\textbf{-43.0}}} & \cellcolor[HTML]{008E98}{\textcolor{white}{\textbf{-34.3}}} &  & \cellcolor[HTML]{006892}{\textcolor{white}{\textbf{-24.4}}} & \cellcolor[HTML]{194F86}{\textcolor{white}{\textbf{-18.4}}} & \cellcolor[HTML]{393A79}{\textcolor{white}{\textbf{-13.5}}}\\
 & -4 & \cellcolor[HTML]{D1E02E}{\textcolor{white}{\textbf{-68.4}}} & \cellcolor[HTML]{A8DA42}{\textcolor{white}{\textbf{-63.8}}} & \cellcolor[HTML]{5ECE63}{\textcolor{white}{\textbf{-56.8}}} &  & \cellcolor[HTML]{00C07B}{\textcolor{white}{\textbf{-50.6}}} & \cellcolor[HTML]{00AC8E}{\textcolor{white}{\textbf{-43.5}}} & \cellcolor[HTML]{009597}{\textcolor{white}{\textbf{-36.5}}} &  & \cellcolor[HTML]{006A92}{\textcolor{white}{\textbf{-24.8}}} & \cellcolor[HTML]{115187}{\textcolor{white}{\textbf{-18.9}}} & \cellcolor[HTML]{3B3778}{\textcolor{white}{\textbf{-13.1}}}\\
\addlinespace
\multirow{2}{*}{$1$} & 0 & \cellcolor[HTML]{8ED64F}{\textcolor{white}{\textbf{-61.3}}} & \cellcolor[HTML]{3BC96D}{\textcolor{white}{\textbf{-54.5}}} & \cellcolor[HTML]{00B587}{\textcolor{white}{\textbf{-46.6}}} &  & \cellcolor[HTML]{00A094}{\textcolor{white}{\textbf{-39.5}}} & \cellcolor[HTML]{008197}{\textcolor{white}{\textbf{-30.8}}} & \cellcolor[HTML]{00608F}{\textcolor{white}{\textbf{-22.5}}} &  & \cellcolor[HTML]{353E7C}{\textcolor{white}{\textbf{-14.5}}} & \cellcolor[HTML]{44276D}{\textcolor{white}{\textbf{-9.7}}} & \cellcolor[HTML]{491462}{\textcolor{white}{\textbf{-6.2}}}\\[-0.33mm]  
\clineThicknessColor{8-8}{1pt}{red}
 & -4 & \cellcolor[HTML]{6DD15E}{\textcolor{white}{\textbf{-58.1}}} & \cellcolor[HTML]{44CA6B}{\textcolor{white}{\textbf{-55.1}}} & \cellcolor[HTML]{00B884}{\textcolor{white}{\textbf{-47.9}}} &  & \cellcolor[HTML]{009896}{\textcolor{white}{\textbf{-37.3}}} & \multicolumn{1}{!{\color{red}\vrule width 1pt}c!{\color{red}\vrule width 1pt}}{\cellcolor[HTML]{008398}{\textcolor{white}{\textbf{-31.4}}}} & \cellcolor[HTML]{006D93}{\textcolor{white}{\textbf{-25.5}}} &  & \cellcolor[HTML]{393A79}{\textcolor{white}{\textbf{-13.6}}} & \cellcolor[HTML]{3F3174}{\textcolor{white}{\textbf{-11.8}}} & \cellcolor[HTML]{471E67}{\textcolor{white}{\textbf{-7.8}}}\\[-0.33mm]  
\clineThicknessColor{8-8}{1pt}{red}
\addlinespace
\multirow{2}{*}{$1/3$} & 0 & \cellcolor[HTML]{00B983}{\textcolor{white}{\textbf{-48.1}}} & \cellcolor[HTML]{00A591}{\textcolor{white}{\textbf{-41.3}}} & \cellcolor[HTML]{009297}{\textcolor{white}{\textbf{-35.3}}} &  & \cellcolor[HTML]{006691}{\textcolor{white}{\textbf{-23.9}}} & \cellcolor[HTML]{055489}{\textcolor{white}{\textbf{-19.5}}} & \cellcolor[HTML]{393A79}{\textcolor{white}{\textbf{-13.6}}} &  & \cellcolor[HTML]{481B65}{\textcolor{white}{\textbf{-7.3}}} & \cellcolor[HTML]{4B045B}{\textcolor{white}{\textbf{-4.5}}} & \cellcolor[HTML]{4B0058}{\textcolor{white}{\textbf{-3.7}}}\\
 & -4 & \cellcolor[HTML]{00AF8C}{\textcolor{white}{\textbf{-44.5}}} & \cellcolor[HTML]{00A293}{\textcolor{white}{\textbf{-40.3}}} & \cellcolor[HTML]{008A98}{\textcolor{white}{\textbf{-33.4}}} &  & \cellcolor[HTML]{005E8E}{\textcolor{white}{\textbf{-22.0}}} & \cellcolor[HTML]{1C4E86}{\textcolor{white}{\textbf{-18.2}}} & \cellcolor[HTML]{3B3778}{\textcolor{white}{\textbf{-13.1}}} &  & \cellcolor[HTML]{4A095D}{\textcolor{white}{\textbf{-5.0}}} & \cellcolor[HTML]{4B045B}{\textcolor{white}{\textbf{-4.6}}} & \cellcolor[HTML]{4B0055}{\textcolor{white}{\textbf{-2.7}}}\\
\end{tabular}}
\caption{\label{tab:PvSR_IntervalScore95_Admin2timesstratum_prevalence}Mean percent change in 95\% interval score of the empirical aggregation model relative to the smooth latent aggregation model, where the response is Admin2 $\times$ stratum prevalence. Yellow-green values are better, while indigo values are worse. Results most representative of the application in Section \ref{sec:application} are outlined in red.}
\end{table}

Tables \ref{tab:PvSR_CRPS_Admin2timesstratum_prevalence} and \ref{tab:PvSR_IntervalScore95_Admin2timesstratum_prevalence} give the Admin2 $\times$ stratum CRPS and 95\% interval relative scores respectively for prevalence, and relative scores for other target quantities and/or at other areal levels are given in the supplement in Section \ref{sec:supplementAdditionalSimStudy}. Overall, the empirical aggregation model performs as good as or better than the smooth latent model for nearly all considered scoring rules and parameters at all areal levels and for all target quantities. In particular, the empirical model performed best when making predictions in areas with small populations, as evidenced by its improvement in CRPS and 95\% interval score being largest when $r_{\tiny \mbox{pop}} = 1/5$.

In addition, the relative performance of the empirical aggregation model improved as the predictive spatial variance decreased relative to the nugget variance, such as when the sample size, $r_{\tiny \mbox{sample}}$, increased, and when the signal to noise ratio, $\varphi$, decreased from 1 to 1/9. This is expected, since under these circumstances, the sources of aggregation error will have large variance relative to the predictive variance of the smooth latent model, particularly fine scale and finite population variability.

Differences between the smooth latent and empirical aggregation models were nearly always larger for the interval score than for CRPS, which is expected since their central predictions are identical for prevalence and burden. The largest improvement in scores from smooth latent to empirical sampling frame models consistently occurred when $r_{\tiny \mbox{samp}} = 3$, $r_{\tiny \mbox{pop}} = 1/5$, and $\varphi = 1/9$. This is because predictive spatial variability is minimized when $r_{\tiny \mbox{samp}} = 3$ (due to having more data in the survey), populations are smallest when $r_{\tiny \mbox{pop}} = 1/5$ (resulting in less fine scale and finite population being averaged out), and fine scale variability is highest compared with smooth spatial variability when $\varphi = 1/9$ (resulting in larger fine scale variability). For example, the smallest (most negative) relative 95\% interval score, $-79.2\%$, occurred under these conditions, and when predicting Admin2 relative prevalence for $\beta_0 = -4$. Hence, the empirical model 95\% interval score was nearly 80\% better than that of the smooth latent due to its ability to account for aggregation uncertainty. The equivalent relative CRPS was $-27.1\%$, indicating a substantial but smaller improvement from the smooth latent to the empirical aggregation model.

Relative scores for parameters most similar to the application indicate smaller but clear performance gain, particularly at the Admin2 and Admin2 $\times$ urban/rural levels. For example, we observe relative 95\% interval (and relative CRPS) scores of -31.4\% (-4.4\%), -4.9\% (-0.5\%), and -0.7\% (-0.1\%) at the Admin2 $\times$ stratum, Admin2, and Admin1 levels respectively when predicting prevalence. Relative scores for relative prevalence are typically more negative than for prevalence and burden, indicating a larger improvement. That is particularly the case for CRPS, since the central prediction for relative prevalence of the three considered aggregation models differ.


\mbox{} \begin{center}
    \begin{sideways}
         \begin{minipage}[t][1.8\textwidth]{1.27\linewidth}
         \includegraphics[width=1\textwidth]{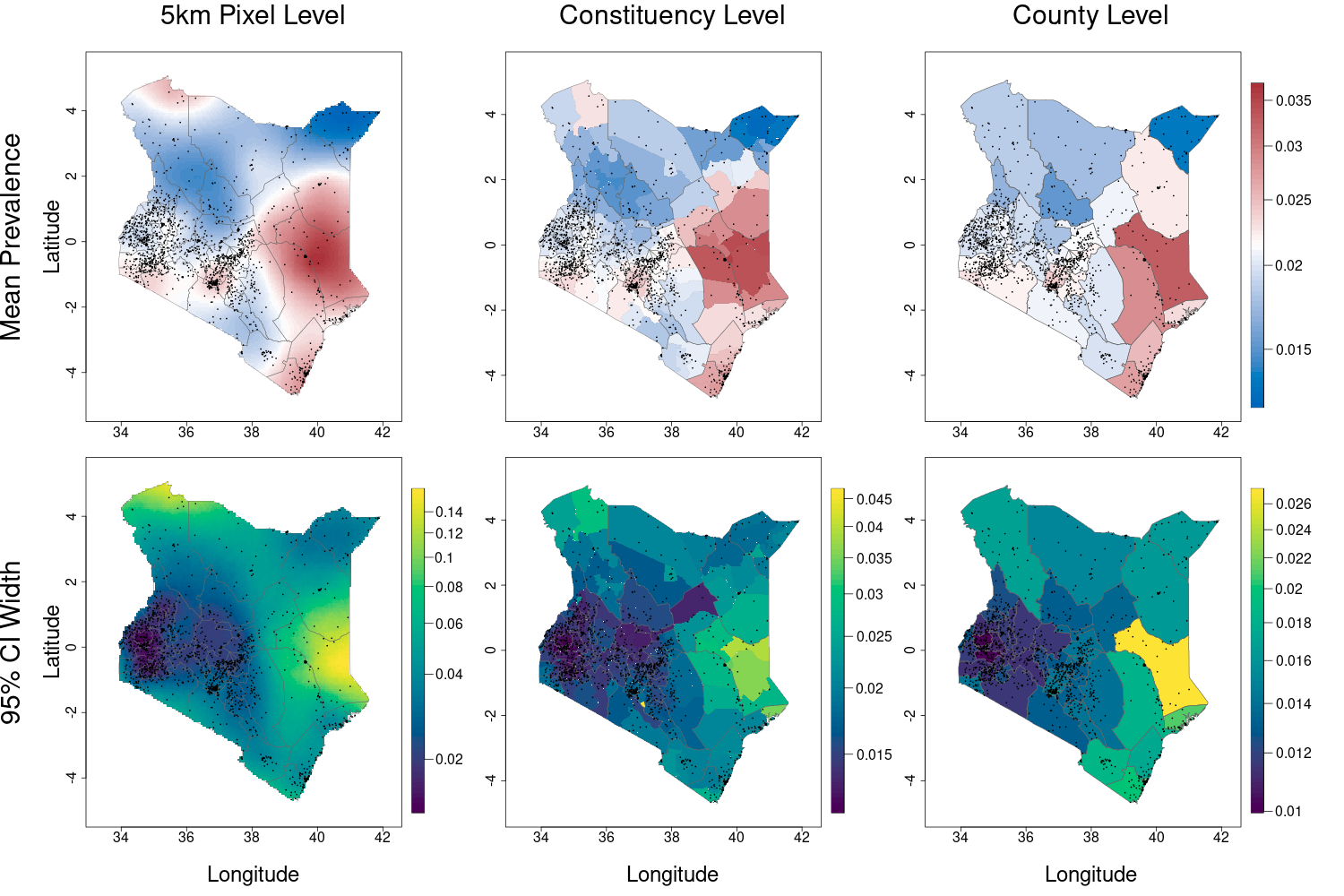}
\captionof{figure}{Central predictions (top row) and 95\% credible interval widths (bottom row) of neonatal mortality rates in Kenya in 2010--2014. Observation locations are plotted as black dots, provinces as thick black lines, and counties as thin gray lines.}
\label{fig:nmrPrevalence}
\end{minipage}
    \end{sideways}
    \end{center}

\section{Application to 2014 KDHS neonatal mortality} \label{sec:application}

We apply the empirical aggregation model, in conjunction with the response model introduced in Section \ref{sec:nuggetEffect}, to 2014 KDHS 2010--2014 NMR data. Unlike for the simulation study, we use an integration grid with 5km rather than 25km resolution for generating areal predictions, since we only need to apply the model to 1 survey rather than 5,400. In addition to generating estimates and 95\% CIs for the prevalence, burden, and relative prevalence at the Admin1, Admin2, and Admin2 $\times$ urban/rural levels when defined, we also generate smooth risk estimates and CIs at the 5km pixel level.

It is important to acknowledge that pixel level smooth risk and smooth burden uncertainties are underreported since they do not account for sources of aggregation error that are especially important at such small spatial scales. While pixel level prediction of the latent and empirical aggregation models is technically possible, their interpretation breaks down at the pixel level due to the fact that EA boundaries can span multiple pixels. In addition, burden estimates are inherently difficult due to the unknown level of uncertainty in the population totals, and the challenge of validation at the areal level when using DHS data since reported DHS survey weights are normalized.

The response model parameter central estimates and other summary statistics are given in Table \ref{tab:riskModTable}.  Central estimates of prevalence along with CIs for EAs are given at the 5km pixel, Admin2, and Admin1 level in Figure \ref{fig:nmrPrevalence}, whereas predictions and CIs for burden and relative prevalence are given in Figure \ref{fig:nmrBurden} and \ref{fig:nmrRelativePrevalence} respectively in the supplement in Section \ref{sec:supplementAdditionalApplication}.

Table \ref{tab:riskModTable} gives little evidence of an urban effect, and the estimated nugget variance is 0.442, which is nearly twice the spatial variance of 0.238. The estimated spatial range of 480 km is somewhat high compared to the diameter of the spatial domain, which is approximately 1,169 km. The long spatial range indicates a large degree of spatial smoothing and a decrease in the spatial variance in the predictive distribution due to increased `borrowing of strength' from observations across long distances.

\begin{table}
\centering
\begin{tabular}{rrrrrr}
 & Est & SD & Q$_{0.025}$ & Q$_{0.5}$ & Q$_{0.975}$ \\ 
\midrule
  Intercept & -4.13 & 0.25 & -4.40 & -4.12 & -3.86 \\
  Urban Effect & -0.00 & 0.11 & -0.15 & -0.00 & 0.14 \\
  Total Var & 0.64 & 0.16 & 0.45 & 0.61 & 0.96 \\
  Spatial Var & 0.20 & 0.16 & 0.06 & 0.16 & 0.56 \\
  Error Var & 0.44 & 0.09 & 0.29 & 0.45 & 0.56 \\
  Total SD & 0.79 & 0.10 & 0.67 & 0.78 & 0.98 \\
  Spatial SD & 0.42 & 0.16 & 0.24 & 0.40 & 0.75 \\
  Error SD & 0.66 & 0.07 & 0.54 & 0.67 & 0.76 \\
  Spatial Range & 411 & 183 & 205 & 356 & 737 \\
\bottomrule
\end{tabular}
\caption{Summary statistics for response model parameters when fit to NMR data from the 2014 KDHS for 2010-2014. Q$_{0.025}$, Q$_{0.5}$, and Q$_{0.975}$ denote the 2.5\textsuperscript{th}, 50\textsuperscript{th}, and 97.5\textsuperscript{th} percentiles respectively.}
\label{tab:riskModTable}
\end{table}

\begin{figure}[ht]
\centering
\includegraphics[width=1.01\textwidth]{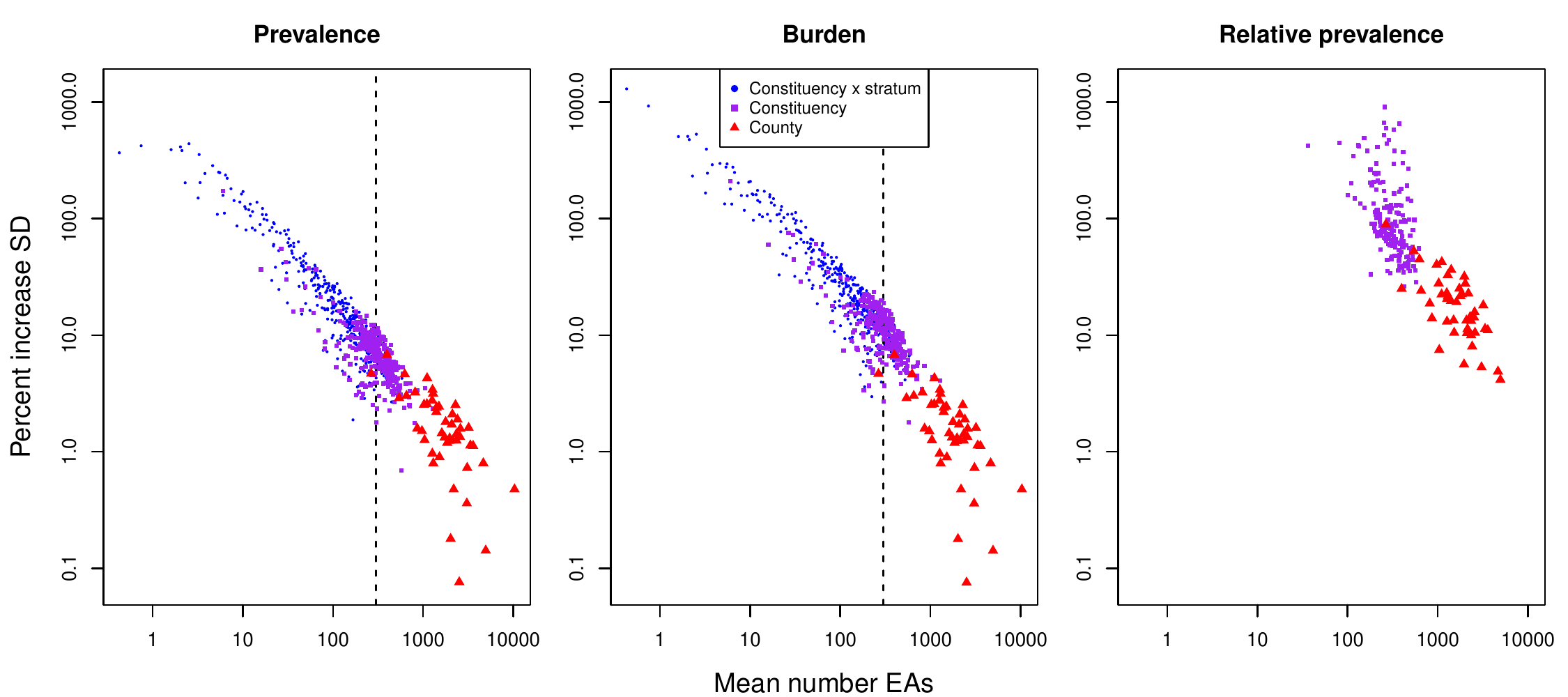}
\caption{Percent increase in posterior standard deviation (SD) of the empirical versus smooth latent models for prevalence (left), burden (middle), and relative prevalence (right). Dashed black lines at 300 mean EAs give a simple threshold for when aggregation error is significant.}
\label{fig:pctIncrease}
\end{figure}

We also find evidence that the sampling frame models make a difference when producing NMR estimates. Figure \ref{fig:pctIncrease} shows how the relative posterior standard deviation (SD) of the empirical to the smooth latent model varies with the number of EAs in an area, calculated as $100\% \times (\widehat{\mbox{SD}}(p_{\tiny \mbox{emp}}(A)) - \widehat{\mbox{SD}}(r_{\tiny \mbox{smooth}}(A))) /$ $\widehat{\mbox{SD}}(r_{\tiny \mbox{smooth}}(A))$ for counties, constituencies, and the urban/rural portion of constituencies. The relative standard deviations in Figure \ref{fig:pctIncrease} are averaged within each area level and each quantity of interest in Table \ref{tab:pctIncrease}, which shows that the empirical model results in 8\% larger uncertainties in prevalence at the Admin2 level, and 31\% larger uncertainties at the Admin2 $\times$ stratum level. The associated relative uncertainties tend to be larger for burden and for relative prevalence.

\begin{table}
\centering
\begin{tabular}{rrrr}
 & \multicolumn{3}{c}{\textit{\textbf{Mean percent increase SD}}}\\
\textbf{Area level} & \textbf{Prevalence} & \textbf{Burden} & \textbf{Relative prevalence} \\ 
\midrule
Constituency $\times$ stratum & 31 & 44 & -- \\ 
  Constituency & 8 & 13 & 132 \\ 
  County & 2 & 2 & 21 \\ 
\bottomrule
\end{tabular}
\caption{Mean percent increase in posterior standard deviation (SD) of the empirical versus smooth latent aggregation models for prevalence, burden, and relative prevalence and for each considered areal level.}
\label{tab:pctIncrease}
\end{table}

As in the simulation study, we find that the relative SD increases as the size of the area, in terms of the number of EAs, decreases. We find the relationship to be approximately linear on a log-log scale for both prevalence and burden, although the relationship is less clear for relative prevalence. Areas with fewer than 300 EAs in this application are especially impacted by the sampling frame model, with posterior SDs that are typically 10 percent higher or more for the empirical aggregation model as opposed to the smooth latent aggregation model. We therefore recommend using the empirical aggregation model for areas finer than the Admin2 level, and for areas with fewer than 300 EAs for this application when estimating prevalence or burden, and we recommend using the empirical aggregation model when estimating relative prevalence even at the Admin1 level for the considered application. The sampling frame model used matters less, however, when estimating prevalence and burden at the Admin1 level in this case.

The urban and rural parts of constituencies with a very small number of average EAs tend to have very high relative SD. For example several Admin2 $\times$ stratum areas have relative SDs of over 500\% when predicting prevalence, and one such area has a relative SD of over 1000\% when predicting burden. This highlights just how much of a difference the sampling frame model can make in some areas, regardless of their average behavior. Moreover, this also indicates that the smooth latent model would likely produce estimates that drastically underestimate uncertainty in very small areas, such as those with fewer than approximately 30 EAs in this application.

\section{Discussion and conclusions} \label{sec:discussion}

Typical geostatistical workflows for spatial aggregation with respect to a population distribution involve the ad-hoc aggregation of point-referenced predictions to the areal level. We propose including a \textit{sampling frame model} to such workflows. The combination of a \textit{response model} for the data and a sampling frame model that incorporates uncertainty in the distribution of the population results in what we call a \textit{spatial aggregation model}. By explicitly incorporating a sampling frame model that includes uncertainty about the population, aggregation models can account for three major sources of aggregation error: aggregation weights, fine scale variation, and finite population variation. Including a sampling frame model also makes more transparent which types of aggregation error are accounted for.

The three main considered sampling frame models are the smooth latent model, the latent model, and the empirical model, which is the model that we propose. The smooth latent model integrates out the spatial nugget effect from the point level risk, creating a smoother risk surface that is integrated with respect to population density. The latent and empirical aggregation models, in contrast, explicitly model how the population is distributed among the enumeration areas (EAs) along with EA locations. This allows them to account for fine scale variability due to EA level effects. The empirical model also accounts for finite population variation by modeling individual outcomes. The empirical model is therefore the only considered sampling frame model that fully accounts for all three major sources of aggregation error.

We have shown that aggregation uncertainty can substantially influence prediction uncertainty in some cases, and so the aggregation procedure should be considered carefully. When the nugget effect in the sampling frame model does not correspond well to that of the response model such as in the `gridded risk' sampling frame model, where a single nugget is included at each spatial aggregation grid point, we have shown that aggregation error can cause predictions to lack robustness to the aggregation grid. If all parameters in the sampling frame model correspond well with those of the spatial response model, the predictions are more robust to the choice of the integration grid, and the grid resolution can be reduced, improving computational performance.

The differences in the considered spatial aggregation models highlights the difference between risk and prevalence, where risk is an expected prevalence, and therefore has less uncertainty. In small area estimation we are ultimately interested in information about an existing population, so we advocate for inference on prevalence. Inference on prevalence results in more conservative uncertainty estimates due to the additional variation of prevalence when compared to risk.

A potential benefit of inference on prevalence is that prevalence is observable while risk is not, making models for prevalence potentially able to be validated more directly. Validating estimates of population averages based on point referenced data is an unsolved problem, but aggregation error may be an essential part of the solution, particularly when it comes to predicting averages of small, left out portions of the data.

While we only considered the effects of aggregation uncertainty in the context of geospatial models, it is possible fine scale and finite population variation may be equally or more important for models with areal spatial effects, since reduced flexibility in the spatial effect could lead to relatively more variance in the nugget. Extending the empirical aggregation model to space-time is also possible, although incorporating multiple age bands in space-time could add complication due to needing to incorporate time and age information to the simulated population in the sampling frame model. 

Ultimately, we find that aggregation uncertainty increases as areas decrease in population size. We observe that aggregation uncertainty is sometimes larger for burden than prevalence due to uncertainty in the population total, and it is often even larger when estimating the relative prevalence between two strata within an area. As a result, we recommend using the empirical or latent aggregation models for small areas, particularly areas smaller than the Admin2 level, and the smooth latent aggregation model for larger areas, such as at the Admin1 level. When estimating relative prevalence in the urban versus rural part of an area, we recommend using the empirical aggregation model even at the Admin1 level, or the latent model when estimating risk.

\section{CRediT authorship contribution statement}
\textbf{John Paige:} Conceptualization, Methodology, Software, Formal analysis, Visualization, Writing - original draft, Writing - review and editing. \textbf{Geir-Arne Fuglstad:} Conceptualization, Writing - review and editing, Supervision. \textbf{Andrea Riebler:} Conceptualization, Writing - review and editing, Supervision. \textbf{Jon Wakefield:} Conceptualization, Writing - review and editing, Supervision.

\section{Data and code}
DHS survey data can be freely obtained at \url{https://dhsprogram.com/}, and WorldPop population density data can be freely obtained at \url{https://www.worldpop.org/}. The postprocessed population density data as well as all code directly used to produce all results in this analysis is available on JP's Github at \url{https://github.com/paigejo/SUMMERdata} and \url{https://github.com/paigejo/continuousNugget} respectively. The code also uses functions that will soon be incorporated into the \verb|SUMMER| R package \citep{SUMMER1.2}, available at \url{https://github.com/paigejo/SUMMER}. Kenya administrative area shapefiles are available at GADM's website at \url{https://gadm.org/index.html}.

\section{Acknowledgements}
This work would not be possible without the financial support of the Research Council of Norway or the National Institutes of Health. They played no role in the formulation of this manuscript, however, and the views expressed in this manuscript are solely those of the authors. The authors declare no further conflicts of interest.

\section{Funding}
GF and AR were supported by project number 240873 from the Research Council of Norway. JW was supported by the National Institutes of Health under award R01~AI029168-30.

\FloatBarrier

\bibliography{myBib}

\pagebreak 
\appendix

\makeatletter\@input{crossReferences2.tex}\makeatother
\end{document}




\begin{frontmatter}

  \title{Supplementary Materials for ``Spatial Aggregation with Respect to a Population Distribution''}

\author[NTNUaddress]{John Paige\corref{mycorrespondingauthor}}
\cortext[mycorrespondingauthor]{Corresponding author}
\ead{john.paige@ntnu.edu}

\author[NTNUaddress]{Geir-Arne Fuglstad}
\author[NTNUaddress]{Andrea Riebler}
\author[UWaddress2]{Jon Wakefield}

\address[NTNUaddress]{Department of Mathematical Sciences, NTNU, Trondheim, Norway}
\address[UWaddress2]{Department of Statistics and Biostatistics, University of Washington, Seattle, USA}

\end{frontmatter}

\appendix
\renewcommand{\thesection}{S.\arabic{section}}

\section{Impact of incorrect aggregation weights on MSE}
\label{sec:supplementAggregationWeights}

Here we will examine the effect of using incorrect aggregation weights has on estimates of population averages and totals given a
quantity of interest, $Y \sim \mbox{GP}(\mu(\cdot), C(\cdot, \cdot))$, spatial mean function $\mu(\cdot)$, and spatial covariance function $C(\cdot, \cdot)$. If the estimated (or assumed) population density is $\hat{q}$, the estimated population total (or average) for region $R$ is,
\begin{equation}
Y(R, \hat{q}) = \int_R \hat{q}(\boldsymbol{s}) Y(\boldsymbol{s}) \ \mathrm{d}\boldsymbol{s}, 
\label{eq:aggregationEquation}
\end{equation}
\noindent 
where $\hat{q}$ is normalized to have unit integral in region $R$ when estimating an average.

We derive the mean square error (MSE) of \eqref{eq:aggregationEquation} with respect to the true population average $Y(R, q)$ for true population density $q$ in the supplement in Section \ref{sec:supplementAggregationWeights}. For $\hat{q}$ being the estimated/assumed population density, and $Y$ defined as above, we find the MSE of our estimate for the population total in region $R$ is: 
\begin{align*}
\mbox{MSE}(Y(R, \hat{q})) &= E \left[ \left( \int_R \hat{q}(\boldsymbol{s}) \mu(\boldsymbol{s}) \ \mathrm{d} \boldsymbol{s} - \int_R q(\boldsymbol{s}) \mu(\boldsymbol{s}) \ \mathrm{d} \boldsymbol{s} \right)^2 \right] \\
&= \left( \int_R (\hat{q} - q)(\boldsymbol{s}) \mu(\boldsymbol{s}) \ \mathrm{d} \boldsymbol{s} \right)^2 - \int_R \int_R (\hat{q} - q)(\boldsymbol{s}) (\hat{q} - q)(\boldsymbol{t}) C(\boldsymbol{s} - \boldsymbol{t}) \ \mathrm{d} \boldsymbol{s} \ \mathrm{d} \boldsymbol{t} \\
&= \mbox{Bias}^2 (Y(R, \hat{q}), Y(R, q)) + \mbox{Var}(Y(R, \hat{q} - q)).
\end{align*}

If we assume that $R$ is sufficiently small, so that $Y(\boldsymbol{s}) = \mu(\boldsymbol{s}) + \epsilon$ for $\boldsymbol{s} \in R$, where $\epsilon \sim N(0, \sigma^2)$, and if we make the further assumption that $\mu$ takes two different values, $\mu_1$ for $\boldsymbol{s} \in A$ for $A \subset R$ and $\mu_2$ for $\boldsymbol{s} \in R \setminus A$ where `$\setminus$' denotes the `set minus' operator. We will assume that $\int_A q(\boldsymbol{s}) \ \mathrm{d} \boldsymbol{s} = q_1$, and $\int_R q(\boldsymbol{s}) \ \mathrm{d} \boldsymbol{s}  = T$, so that $\int_{R\setminus A} q(\boldsymbol{s}) \ \mathrm{d} \boldsymbol{s}  = q_2 = T - q_1$. We similarly assume that the integral of $\hat{q}$ over $A$, $R$, and $R \setminus A$ is $\hat{q}_1$, $\hat{T}$, and $\hat{q}_2 = \hat{T} - \hat{q}_1$ respectively. The resulting bias is:
\begin{align*}
\mbox{Bias} (Y(R, \hat{q}), Y(R, q)) &= \int_R (\hat{q} - q)(\boldsymbol{s}) \mu(\boldsymbol{s}) \ \mathrm{d} \boldsymbol{s} \\
&= (\hat{q}_1 - q_1) \mu_1 + (\hat{q}_2 - q_2) \mu_2 \\
&= (\hat{q}_1 - q_1) \mu_1 + (\hat{T} - T + q_1 - \hat{q}_1) \mu_2 \\
&= (\hat{q}_1 - q_1) (\mu_1 - \mu_2) + (\hat{T} - T) \mu_2.
\end{align*}
The variance is:
\begin{align*}
\mbox{Var}(Y(R, \hat{q} - q)) &= \int_R \int_R (\hat{q} - q)(\boldsymbol{s}) (\hat{q} - q)(\boldsymbol{t}) C(\boldsymbol{s} - \boldsymbol{t}) \ \mathrm{d} \boldsymbol{s} \ \mathrm{d} \boldsymbol{t} \\
&= \sigma^2 (\hat{q}_1 q_1 + \hat{q}_1(T - q_1) + (\hat{T} - \hat{q}_1)q_1 + (\hat{T} - \hat{q}_1)(T - q_1)) \\
&= \sigma^2 T \hat{T}.
\end{align*}
Hence, the MSE is:
\begin{align*}
\mbox{MSE}(Y(R, \hat{q})) &= \left( (\hat{q}_1 - q_1) \mu_1 + (\hat{q}_2 - q_2) \mu_2 \right)^2 +  \sigma^2 T \hat{T} \\
&= (\hat{q}_1 - q_1)^2 (\mu_1 - \mu_2)^2 + 2 (\hat{q}_1 - q_1) (\mu_1 - \mu_2) (\hat{T} - T) \mu_2 + (\hat{T} - T)^2 \mu_2^2 +  \sigma^2 T \hat{T}.
\end{align*}
If $\hat{T} = T$, such as in the case where $q$ and $\hat{q}$ are both normalized to 1 (for example, when estimating risk or prevalence), then the MSE simplifies to:
\begin{align*}
\mbox{MSE}(Y(R, \hat{q})) &= (\hat{q}_1 - q_1)^2 (\mu_1 - \mu_2)^2 +  \sigma^2 T^2.
\end{align*}


If we assume that $R$ is sufficiently small, so that $Y(\boldsymbol{s}) = \mu(\boldsymbol{s}) + \epsilon$ for $\boldsymbol{s} \in R$, where $\epsilon \sim N(0, \sigma^2)$, and if we make the further assumption that $\mu$ takes two different values, $\mu_1$ for $\boldsymbol{s} \in A$ for $A \subset R$ and $\mu_2$ for $\boldsymbol{s} \in R \setminus A$ where `$\setminus$' denotes the `set minus' operator. We will assume that $\int_A q(\boldsymbol{s}) \ \mathrm{d} \boldsymbol{s} = q_1$, and $\int_{R\setminus A} q(\boldsymbol{s}) \ \mathrm{d} \boldsymbol{s}  = q_2$. We similarly assume that the integral of $\hat{q}$ over $A$ and $R \setminus A$ is $\hat{q}_1$ and $\hat{q}_2$ respectively. The resulting bias is:
\begin{align*}
\mbox{Bias} (Y(R, \hat{q}), Y(R, q)) &= \int_R (\hat{q} - q)(\boldsymbol{s}) \mu(\boldsymbol{s}) \ \mathrm{d} \boldsymbol{s} \\
&= (\hat{q}_1 - q_1) \mu_1 + (\hat{q}_2 - q_2) \mu_2.
\end{align*}
The variance is:
\begin{align*}
\mbox{Var}(Y(R, \hat{q} - q)) &= \int_R \int_R (\hat{q} - q)(\boldsymbol{s}) (\hat{q} - q)(\boldsymbol{t}) C(\boldsymbol{s} - \boldsymbol{t}) \ \mathrm{d} \boldsymbol{s} \ \mathrm{d} \boldsymbol{t} \\
&= \sigma^2 (\hat{q}_1 q_1 + \hat{q}_1 q_2 +  \hat{q}_2 q_1 + \hat{q}_2 q_2) \\
&= \sigma^2 T \hat{T}
\end{align*}
for $T = q_1 + q_2$ and $\hat{T} = \hat{q}_1 + \hat{q}_2$. Hence, the MSE is:
\begin{align*}
\mbox{MSE}(Y(R, \hat{q})) &= \left( (\hat{q}_1 - q_1) \mu_1 + (\hat{q}_2 - q_2) \mu_2 \right)^2 +  \sigma^2 T \hat{T} \\
&= (\hat{q}_1 - q_1)^2 \mu_1^2 + 2 (\hat{q}_1 - q_1) (\hat{q}_2 - q_2) \mu_1 \mu_2 + (\hat{q}_2 - q_2)^2 \mu_2^2 +  \sigma^2 T \hat{T}.
\end{align*}
If $\hat{q}_1$ and $\hat{q}_2$ are random with $E[\hat{q}_1] = q_1 + b_1$ and $E[\hat{q}_2] = q_2 + b_2$ with $\mbox{Var}(\hat{q}_1) = \mbox{Var}(\hat{q}_2) = \nu^2$, and assuming $\hat{q}_1 \perp \hat{q}_2$, then we have:
\begin{align*}
\mbox{MSE}(Y(R, \hat{q})) &= (b_1^2 + \nu^2) \mu_1^2 + 2 b_1 b_2 \mu_1 \mu_2 + (b_2^2 + \nu^2) \mu_2^2 + \sigma^2 T(T + b_1 + b_2).
\end{align*}
If $\hat{T} = T$, such as in the case where $q$ and $\hat{q}$ are both normalized to 1 (for example, when estimating risk or prevalence), then
\begin{align*}
E[(\hat{q}_1 - q_1) (\hat{q}_2 - q_2) \mid \hat{T} = T] &= E[(\hat{q}_1 - q_1) (T - \hat{q}_1 - T + q_1) \mid \hat{T} = T] \\
&= E[-(\hat{q}_1 - q_1)^2 \mid \hat{T} = T]  \leq 0.
\end{align*}
If $\mu_1, \mu_2 \geq 0$ the cross term of the MSE is then nonpositive, as would be the case if $\mu$ represented a risk or pollution levels.

\section{Integration grid generation and urbanicity classification}
\label{sec:supplementIntegrationGrid}

To generate an integration grid over Kenya, we first project the approximate latitudinal and longitudinal extent of Kenya using a projection with European Petroleum Survey Group (EPSG) code 21097, which is accurate to 6m in Kenya \citep{kenyaProjection}. Once the extent is projected into easting/northing (in kilometers) we construct a grid with uniform resolution under the projected coordinate system, removing points outside of Kenya from the integration grid. If certain areas of interest contain no integration points, custom integration points are added at the centroids of those areas. This add flexibility to the integration grid resolution, in particular in the presence of small areas of interest that may not be well represented at a given resolution otherwise.

We first construct a fine scale grid with 1 km resolution for the purpose of estimating urban and rural population at the admin2 level. We assume, as is the case for Kenya, that estimates of general population totals are available at the admin1 level and for the urban and rural portions of each Admin1 area. In the case of Kenya, this is available in the 2009 Kenya Population and Housing Census. Given all spatial locations in the integration grid in a given admin1 area, say $\boldsymbol{s}_1, \ldots, \boldsymbol{s}_L$, with general population densities, $\rho(\boldsymbol{s}_1), \ldots, \rho(\boldsymbol{s}_L)$, we first set a threshold, $K$, so that, 
$$K = \argmin_{k} \frac{\sum_{l : \rho(\boldsymbol{s}_l) \geq k} \rho(\boldsymbol{s}_l)}{\sum_{l} \rho(\boldsymbol{s}_l)} \geq P^{\tiny \mbox{RUR}}, $$
\noindent
where $P^{\tiny \mbox{RUR}}$ is the proportion of general population that is rural in the Admin1 area of interest. We then classify location $\boldsymbol{s}_l$ as urban if $q(\boldsymbol{s}_l) \geq K$ for $l=1, \ldots, L$.

All of this is performed in the SUMMER package in R \citep{SUMMER1.2} via the \texttt{makePopIntegrationTab} function, using population density data and shapefiles provided on github at \url{https://github.com/paigejo/SUMMERdata}.

\section{Relationship between the aggregation models} \label{sec:supplementRelationshipBetweenAgg}

\subsection{Shared expectations}
We show in this section that the gridded, empirical, latent, and smooth latent models all have the same central predictions. This is because $E_{\mathbf{Z}}[p_{\tiny \mbox{emp}}(A)] = r_{\tiny \mbox{latent}}(A)$, where $\mathbf{Z} = \{Z_i \}_{i=1}^M$ and the subscript of the expectation denotes what is expected over. Further, $E_{\mathbf{Z}, \mathbf{N}, \boldsymbol{\epsilon}}[p_{\tiny \mbox{emp}}(A)] = E_{\mathbf{N}, \boldsymbol{\epsilon}}[r_{\tiny \mbox{latent}}(A)] = r_{\tiny \mbox{smooth}}(A) = E_{\boldsymbol{\epsilon}}[r_{\tiny \mbox{grid}}(A)]$, where $\mathbf{N} = \{N_i \}_{i=1}^M$ and $\boldsymbol{\epsilon} = \{\epsilon_i \}_{i=1}^M$. For similar reasons, the expectation of the four sampling frame models considered are also the same when applied to burden.

While it is clear that the smooth latent and gridded models use \eqref{eq:areal} to estimate areal smooth risk and smooth burden choosing a continuously indexed population density $q$ as the spatial density, it is worth noting that the empirical and latent models can also be viewed as using \eqref{eq:areal}, choosing the integrand, $r(\boldsymbol{s})$, to be either the EA level prevalence or risk respectively. Rather than integrating $r(\boldsymbol{s})$ over a continuously indexed spatial density, the empirical and latent models choose $Q(\boldsymbol{s})$ to be a point process for the EA locations and integrate with respect to this point process. Hence, the empirical, latent, and smooth latent sampling frame models are all closely related.

For the latent aggregation model, the areal estimates can be shown to be an expectation over part of the sampling frame model of the empirical areal estimates: 
\begin{align*}
E_{\boldsymbol{Z}}  \left [ p_{\tiny \mbox{emp}}(A) \right ] &= E_{\boldsymbol{Z}}  \left [ \sum_{i: \boldsymbol{s}_i \in A} \frac{N_i}{N}  \times \frac{Z_i}{N_i} \right ] \\
&= \sum_{i: \boldsymbol{s}_i \in A} \frac{N_i}{N}  \times E_{\boldsymbol{Z}}  \left [ \frac{Z_i}{N_i} \right ] \\
&= \sum_{i: \boldsymbol{s}_i \in A} \frac{N_i}{N} \times  r_{i} \\
&= r_{\tiny \mbox{latent}}(A).
\end{align*}
Hence, the tower property of expectations implies $E \left [ p_{\tiny \mbox{emp}}(A) \right ] = E[r_{\tiny \mbox{latent}}(A)]$. Further, 
\begin{align*}
E_{\boldsymbol{Z},\boldsymbol{N}, \boldsymbol{\epsilon}, \{\boldsymbol{s}_i\}_{i=1}^M}  \left [ p(A) \right ] &= E_{\boldsymbol{Z},\boldsymbol{N}, \boldsymbol{\epsilon}, \{\boldsymbol{s}_i\}_{i=1}^M}  \left [ \sum_{i: \boldsymbol{s}_i \in A} \frac{N_i}{N}  \times \frac{Z_i}{N_i} \right ] \\
&= E_{\boldsymbol{N}, \{\boldsymbol{s}_i\}_{i=1}^M}  \left [ \sum_{i: \boldsymbol{s}_i \in A} \frac{N_i}{N}  \times E_{\boldsymbol{Z}, \boldsymbol{\epsilon}}  \left [ \frac{Z_i}{N_i} \right ] \right ] \\
&= E_{\{\boldsymbol{s}_i\}_{i=1}^M}  \left [ \sum_{i: \boldsymbol{s}_i \in A} E_{\boldsymbol{N}}  \left [ \frac{N_i}{N} \right ] \times  E_{\boldsymbol{\epsilon}}[r_{i}] \right ] \\
&= E_{\{\boldsymbol{s}_i\}_{i=1}^M} \left [ \sum_{i: \boldsymbol{s}_i \in A} Q_A \times  r_{\tiny \mbox{smooth}}(\boldsymbol{s}_i) \right ] \\
&= \int_A q(\boldsymbol{s}) Q_A \times r_{\tiny \mbox{smooth}}(\boldsymbol{s}) \ \mathrm{d} \boldsymbol{s}, \\
&= r_{\tiny \mbox{smooth}}(A),
\end{align*}
where $q$ is proportional to the population density and $Q_A$ is the expected proportion of the total target population in $A$ contained in a single EA in $A$, so that $\int_A q(\boldsymbol{s}) Q_A \ \mathrm{d} \boldsymbol{s} = 1$. It is easy to see that $E[r_{\tiny \mbox{grid}}(A)] = E[r_{\tiny \mbox{smooth}}(A)]$. Hence, $E \left [ p_{\tiny \mbox{emp}}(A) \right ] = E[r_{\tiny \mbox{latent}}(A)] = E[r_{\tiny \mbox{grid}}(A)] = E[r_{\tiny \mbox{smooth}}(A)]$. A similar argument shows that $E \left [ b_{\tiny \mbox{emp}}(A) \right ] = E[b_{\tiny \mbox{latent}}(A)] = E[b_{\tiny \mbox{grid}}(A)] = E[b_{\tiny \mbox{smooth}}(A)]$.

When producing estimates for an administrative area stratified by urban/rural, one can take a weighted average (or a sum, in the case of burden) of the estimates in the urban and rural parts of that area, weighted by the target population totals in the urban and rural parts of the area.

\subsection{Relative variances}

The law of total variance and some algebra shows that the variance of the empirical aggregation model predictions is necessarily greater than or equal to the variance of the smooth latent predictions when it comes to prevalence and burden. Assuming there is a nonzero target population in area $A$, we find:
\begin{align*}
\mbox{Var}(p_{\tiny \mbox{emp}}(A)) &= E[\mbox{Var}_{\boldsymbol{Z},\boldsymbol{N}, \boldsymbol{\epsilon}, \{\boldsymbol{s}_i\}_{i=1}^M}(p_{\tiny \mbox{emp}}(A))] + \mbox{Var}(E_{\boldsymbol{Z},\boldsymbol{N}, \boldsymbol{\epsilon}, \{\boldsymbol{s}_i\}_{i=1}^M}[p_{\tiny \mbox{emp}}(A)]) \\
&= E[\mbox{Var}_{\boldsymbol{Z},\boldsymbol{N}, \boldsymbol{\epsilon}, \{\boldsymbol{s}_i\}_{i=1}^M}(p_{\tiny \mbox{emp}}(A))] + \mbox{Var}(r_{\tiny \mbox{smooth}}(A)).
\end{align*}
With similar arguments it is trivial to show, 
\begin{align*}
\mbox{Var}(p_{\tiny \mbox{emp}}(A)) &\geq \mbox{Var}(r_{\tiny \mbox{latent}}(A)) \geq \mbox{Var}(r_{\tiny \mbox{smooth}}(A)), \quad \text{and} \\
\mbox{Var}(b_{\tiny \mbox{emp}}(A)) &\geq \mbox{Var}(b_{\tiny \mbox{latent}}(A)) \geq \mbox{Var}(b_{\tiny \mbox{smooth}}(A)). 
\end{align*}
While this argument does not apply directly to the variance of relative prevalence compared with the variance of the relative smooth risk since these are ratios, we find evidence in Section \ref{sec:application} that the variance of relative prevalence is not only larger than the variance of smooth risk, but larger by more a greater factor than for burden and prevalence/risk, provided the relative prevalence and relative burden are defined.

\clearpage

\section{Additional grid resolution test results}
\label{sec:supplementAdditionalGridRes}

\begin{table}[h] \centering
\begin{tabular}{llcrrrr}
\toprule[0.09 em]
& \textbf{Model} & \textbf{Units} & \textbf{200m} & \textbf{1km} & \textbf{5km} & \textbf{25km} \\ 
  \hline
50\% CI width & Empirical & (per 1,000) & 3.3 & 3.3 & 3.3 & 3.3 \\ 
& Latent && 2.7 & 2.7 & 2.7 & 2.7 \\ 
& Smooth latent && 2.5 & 2.5 & 2.5 & 2.6 \\ 
& Gridded && 2.8 & 6.3 & 16.3 & 17.5 \\ 
\addlinespace[0.3em]
50\% CI coverage & Empirical & (Percent) & 52 & 52 & 52 & 52 \\ 
& Latent && 44 & 44 & 44 & 45 \\ 
& Smooth latent && 42 & 42 & 42 & 42 \\ 
& Gridded && 46 & 76 & 98 & 99 \\ 
\bottomrule[0.09 em]
\end{tabular}
\caption{50\% credible interval (CI) widths in neonatals per thousand, and empirical coverages in percent for considered sampling frame models as a function of aggregation grid resolution.}
\label{tab:gridResTest50}
\end{table}

\clearpage

\section{Additional simulation study results}
\label{sec:supplementAdditionalSimStudy}

\makeatletter
\def\clineThicknessColor#1#2#3{\@ClineThicknessColor#1\@nil{#2}{#3}}
\def\@ClineThicknessColor#1-#2\@nil#3#4{%
    \omit
    \@multicnt#1%
    \advance\@multispan\m@ne
    \ifnum\@multicnt=\@ne\@firstofone{&\omit}\fi
    \@multicnt#2%
    \advance\@multicnt-#1%
    \advance\@multispan\@ne
    {\color{#4}%
    \leaders\hrule\@height#3\hfill}%
    \cr}
\makeatother

\FloatBarrier
\subsection{Admin2 $\times$ stratum burden}
\FloatBarrier

\begin{table}[h]
\centering
\resizebox{\linewidth}{!}{
\begin{tabular}[t]{c@{\phantom{\tiny a}}c|ccc@{\phantom{a}}p{0em}ccc@{\phantom{a}}p{0em}ccc}
  & $r_{\tiny \mbox{pop}}$ & \multicolumn{3}{c}{$1/5$} &   & \multicolumn{3}{c}{$1$} &   & \multicolumn{3}{c}{$5$} \\[-0.5ex]
$r_{\tiny \mbox{samp}}$ & \diagbox[width=.3in, height=.3in, innerleftsep=.04in, innerrightsep=-.02in]{$\beta_0$}{$\varphi$} & $1/9$ & $1/3$ & $1$ &   & $1/9$ & $1/3$ & $1$ &   & $1/9$ & $1/3$ & $1$ \\
\hline
\multirow{2}{*}{$3$} & 0 & \cellcolor[HTML]{D1E02E}{\textcolor{white}{\textbf{-19.5}}} & \cellcolor[HTML]{89D551}{\textcolor{white}{\textbf{-17.2}}} & \cellcolor[HTML]{50CC68}{\textcolor{white}{\textbf{-15.8}}} &  & \cellcolor[HTML]{00A093}{\textcolor{white}{\textbf{-11.0}}} & \cellcolor[HTML]{008097}{\textcolor{white}{\textbf{-8.2}}} & \cellcolor[HTML]{006691}{\textcolor{white}{\textbf{-6.2}}} &  & \cellcolor[HTML]{383B7A}{\textcolor{white}{\textbf{-3.2}}} & \cellcolor[HTML]{46226A}{\textcolor{white}{\textbf{-1.7}}} & \cellcolor[HTML]{481C66}{\textcolor{white}{\textbf{-1.3}}}\\
 & -4 & \cellcolor[HTML]{00A193}{\textcolor{white}{\textbf{-11.0}}} & \cellcolor[HTML]{008297}{\textcolor{white}{\textbf{-8.4}}} & \cellcolor[HTML]{00618F}{\textcolor{white}{\textbf{-5.9}}} &  & \cellcolor[HTML]{3A3878}{\textcolor{white}{\textbf{-3.0}}} & \cellcolor[HTML]{46226A}{\textcolor{white}{\textbf{-1.6}}} & \cellcolor[HTML]{491763}{\textcolor{white}{\textbf{-1.1}}} &  & \cellcolor[HTML]{4A095D}{\textcolor{white}{\textbf{-0.6}}} & \cellcolor[HTML]{4B0058}{\textcolor{white}{\textbf{-0.2}}} & \cellcolor[HTML]{4B0057}{\textcolor{white}{\textbf{-0.1}}}\\
\addlinespace
\multirow{2}{*}{$1$} & 0 & \cellcolor[HTML]{73D25B}{\textcolor{white}{\textbf{-16.5}}} & \cellcolor[HTML]{00BE7D}{\textcolor{white}{\textbf{-14.0}}} & \cellcolor[HTML]{00AD8D}{\textcolor{white}{\textbf{-12.2}}} &  & \cellcolor[HTML]{006F94}{\textcolor{white}{\textbf{-6.9}}} & \cellcolor[HTML]{005589}{\textcolor{white}{\textbf{-4.9}}} & \cellcolor[HTML]{353E7C}{\textcolor{white}{\textbf{-3.4}}} &  & \cellcolor[HTML]{481C66}{\textcolor{white}{\textbf{-1.4}}} & \cellcolor[HTML]{4A0C5E}{\textcolor{white}{\textbf{-0.7}}} & \cellcolor[HTML]{4A095D}{\textcolor{white}{\textbf{-0.6}}}\\[-0.33mm]  
\clineThicknessColor{8-8}{1pt}{red}
 & -4 & \cellcolor[HTML]{006C93}{\textcolor{white}{\textbf{-6.7}}} & \cellcolor[HTML]{00598B}{\textcolor{white}{\textbf{-5.2}}} & \cellcolor[HTML]{33417E}{\textcolor{white}{\textbf{-3.6}}} &  & \cellcolor[HTML]{471F68}{\textcolor{white}{\textbf{-1.5}}} & \multicolumn{1}{!{\color{red}\vrule width 1pt}c!{\color{red}\vrule width 1pt}}{\cellcolor[HTML]{491261}{\textcolor{white}{\textbf{-0.9}}}} & \cellcolor[HTML]{4A0C5E}{\textcolor{white}{\textbf{-0.7}}} &  & \cellcolor[HTML]{4B0057}{\textcolor{white}{\textbf{-0.1}}} & \cellcolor[HTML]{4B0057}{\textcolor{white}{\textbf{-0.1}}} & \cellcolor[HTML]{4B0057}{\textcolor{white}{\textbf{-0.1}}}\\[-0.33mm]  
\clineThicknessColor{8-8}{1pt}{red}
\addlinespace
\multirow{2}{*}{$1/3$} & 0 & \cellcolor[HTML]{00AD8D}{\textcolor{white}{\textbf{-12.1}}} & \cellcolor[HTML]{009297}{\textcolor{white}{\textbf{-9.7}}} & \cellcolor[HTML]{008598}{\textcolor{white}{\textbf{-8.6}}} &  & \cellcolor[HTML]{353E7C}{\textcolor{white}{\textbf{-3.4}}} & \cellcolor[HTML]{3F3174}{\textcolor{white}{\textbf{-2.6}}} & \cellcolor[HTML]{44296E}{\textcolor{white}{\textbf{-2.0}}} &  & \cellcolor[HTML]{4A1060}{\textcolor{white}{\textbf{-0.8}}} & \cellcolor[HTML]{4B0058}{\textcolor{white}{\textbf{-0.2}}} & \cellcolor[HTML]{4B0058}{\textcolor{white}{\textbf{-0.2}}}\\
 & -4 & \cellcolor[HTML]{3B3778}{\textcolor{white}{\textbf{-2.9}}} & \cellcolor[HTML]{3F3174}{\textcolor{white}{\textbf{-2.6}}} & \cellcolor[HTML]{45266C}{\textcolor{white}{\textbf{-1.9}}} &  & \cellcolor[HTML]{4B0057}{\textcolor{white}{\textbf{0.0}}} & \cellcolor[HTML]{4B0059}{\textcolor{white}{\textbf{-0.3}}} & \cellcolor[HTML]{4B0059}{\textcolor{white}{\textbf{-0.3}}} &  & \cellcolor[HTML]{4B0055}{\textcolor{white}{\textbf{0.1}}} & \cellcolor[HTML]{4B0057}{\textcolor{white}{\textbf{-0.1}}} & \cellcolor[HTML]{4B0057}{\textcolor{white}{\textbf{0.0}}}\\
\end{tabular}}
\vspace{-.3cm}
\caption{\label{tab:PvSR_CRPS_Admin2timesstratum_burden}Mean percent change in CRPS of the empirical aggregation model relative to the smooth latent aggregation model, where the response is Admin2 $\times$ stratum burden. Yellow-green values are better, while indigo values are worse. Results most representative of the application in Section \ref{sec:application} are outlined in red.}
\end{table}

\begin{table}[h]
\centering
\resizebox{\linewidth}{!}{
\begin{tabular}[t]{c@{\phantom{\tiny a}}c|ccc@{\phantom{a}}p{0em}ccc@{\phantom{a}}p{0em}ccc}
  & $r_{\tiny \mbox{pop}}$ & \multicolumn{3}{c}{$1/5$} &   & \multicolumn{3}{c}{$1$} &   & \multicolumn{3}{c}{$5$} \\[-0.5ex]
$r_{\tiny \mbox{samp}}$ & \diagbox[width=.3in, height=.3in, innerleftsep=.04in, innerrightsep=-.02in]{$\beta_0$}{$\varphi$} & $1/9$ & $1/3$ & $1$ &   & $1/9$ & $1/3$ & $1$ &   & $1/9$ & $1/3$ & $1$ \\
\hline
\multirow{2}{*}{$3$} & 0 & \cellcolor[HTML]{D1E02E}{\textcolor{white}{\textbf{-75.7}}} & \cellcolor[HTML]{BBDD38}{\textcolor{white}{\textbf{-72.6}}} & \cellcolor[HTML]{A8DA42}{\textcolor{white}{\textbf{-70.3}}} &  & \cellcolor[HTML]{3BC96D}{\textcolor{white}{\textbf{-59.7}}} & \cellcolor[HTML]{00B686}{\textcolor{white}{\textbf{-51.0}}} & \cellcolor[HTML]{00A193}{\textcolor{white}{\textbf{-42.9}}} &  & \cellcolor[HTML]{006C93}{\textcolor{white}{\textbf{-25.7}}} & \cellcolor[HTML]{2D4580}{\textcolor{white}{\textbf{-15.0}}} & \cellcolor[HTML]{373C7B}{\textcolor{white}{\textbf{-12.7}}}\\
 & -4 & \cellcolor[HTML]{24C771}{\textcolor{white}{\textbf{-58.3}}} & \cellcolor[HTML]{00B488}{\textcolor{white}{\textbf{-50.1}}} & \cellcolor[HTML]{009896}{\textcolor{white}{\textbf{-39.6}}} &  & \cellcolor[HTML]{006390}{\textcolor{white}{\textbf{-23.2}}} & \cellcolor[HTML]{2D4580}{\textcolor{white}{\textbf{-14.8}}} & \cellcolor[HTML]{3E3475}{\textcolor{white}{\textbf{-10.6}}} &  & \cellcolor[HTML]{471E67}{\textcolor{white}{\textbf{-5.4}}} & \cellcolor[HTML]{4A095D}{\textcolor{white}{\textbf{-2.1}}} & \cellcolor[HTML]{4B025A}{\textcolor{white}{\textbf{-1.1}}}\\
\addlinespace
\multirow{2}{*}{$1$} & 0 & \cellcolor[HTML]{AFDB3E}{\textcolor{white}{\textbf{-71.1}}} & \cellcolor[HTML]{89D551}{\textcolor{white}{\textbf{-66.6}}} & \cellcolor[HTML]{64CF61}{\textcolor{white}{\textbf{-62.7}}} &  & \cellcolor[HTML]{00A790}{\textcolor{white}{\textbf{-44.8}}} & \cellcolor[HTML]{008E98}{\textcolor{white}{\textbf{-36.2}}} & \cellcolor[HTML]{007595}{\textcolor{white}{\textbf{-28.5}}} &  & \cellcolor[HTML]{3A3878}{\textcolor{white}{\textbf{-11.7}}} & \cellcolor[HTML]{472069}{\textcolor{white}{\textbf{-6.0}}} & \cellcolor[HTML]{481C66}{\textcolor{white}{\textbf{-5.0}}}\\[-0.33mm]  
\clineThicknessColor{8-8}{1pt}{red}
 & -4 & \cellcolor[HTML]{009A95}{\textcolor{white}{\textbf{-40.4}}} & \cellcolor[HTML]{008998}{\textcolor{white}{\textbf{-34.7}}} & \cellcolor[HTML]{006C93}{\textcolor{white}{\textbf{-25.7}}} &  & \cellcolor[HTML]{3F3174}{\textcolor{white}{\textbf{-9.9}}} & \multicolumn{1}{!{\color{red}\vrule width 1pt}c!{\color{red}\vrule width 1pt}}{\cellcolor[HTML]{45256B}{\textcolor{white}{\textbf{-7.0}}}} & \cellcolor[HTML]{481C66}{\textcolor{white}{\textbf{-5.2}}} &  & \cellcolor[HTML]{4B0058}{\textcolor{white}{\textbf{-0.6}}} & \cellcolor[HTML]{4B025A}{\textcolor{white}{\textbf{-1.1}}} & \cellcolor[HTML]{4B0059}{\textcolor{white}{\textbf{-0.8}}}\\[-0.33mm]  
\clineThicknessColor{8-8}{1pt}{red}
\addlinespace
\multirow{2}{*}{$1/3$} & 0 & \cellcolor[HTML]{4CCB69}{\textcolor{white}{\textbf{-60.9}}} & \cellcolor[HTML]{00BF7C}{\textcolor{white}{\textbf{-54.9}}} & \cellcolor[HTML]{00B884}{\textcolor{white}{\textbf{-51.9}}} &  & \cellcolor[HTML]{006C93}{\textcolor{white}{\textbf{-25.6}}} & \cellcolor[HTML]{005A8C}{\textcolor{white}{\textbf{-20.7}}} & \cellcolor[HTML]{1C4E86}{\textcolor{white}{\textbf{-17.5}}} &  & \cellcolor[HTML]{481C66}{\textcolor{white}{\textbf{-4.9}}} & \cellcolor[HTML]{4A0C5E}{\textcolor{white}{\textbf{-2.4}}} & \cellcolor[HTML]{4A065C}{\textcolor{white}{\textbf{-1.7}}}\\
 & -4 & \cellcolor[HTML]{00568A}{\textcolor{white}{\textbf{-19.4}}} & \cellcolor[HTML]{1F4D85}{\textcolor{white}{\textbf{-17.1}}} & \cellcolor[HTML]{343F7D}{\textcolor{white}{\textbf{-13.5}}} &  & \cellcolor[HTML]{4B045B}{\textcolor{white}{\textbf{-1.4}}} & \cellcolor[HTML]{4B025A}{\textcolor{white}{\textbf{-1.1}}} & \cellcolor[HTML]{4A065C}{\textcolor{white}{\textbf{-1.9}}} &  & \cellcolor[HTML]{4B0055}{\textcolor{white}{\textbf{0.6}}} & \cellcolor[HTML]{4B0057}{\textcolor{white}{\textbf{-0.3}}} & \cellcolor[HTML]{4B0057}{\textcolor{white}{\textbf{0.0}}}\\
\end{tabular}}
\vspace{-.3cm}
\caption{\label{tab:PvSR_IntervalScore95_Admin2timesstratum_burden}Mean percent change in 95\% interval score of the empirical aggregation model relative to the smooth latent aggregation model, where the response is Admin2 $\times$ stratum burden. Yellow-green values are better, while indigo values are worse. Results most representative of the application in Section \ref{sec:application} are outlined in red.}
\end{table}

\FloatBarrier
\clearpage
\subsection{Admin2 prevalence}
\FloatBarrier

\begin{table}[h]
\centering
\resizebox{.96\linewidth}{!}{
\begin{tabular}[t]{c@{\phantom{\tiny a}}c|ccc@{\phantom{a}}p{0em}ccc@{\phantom{a}}p{0em}ccc}
  & $r_{\tiny \mbox{pop}}$ & \multicolumn{3}{c}{$1/5$} &   & \multicolumn{3}{c}{$1$} &   & \multicolumn{3}{c}{$5$} \\[-0.5ex]
$r_{\tiny \mbox{samp}}$ & \diagbox[width=.3in, height=.3in, innerleftsep=.04in, innerrightsep=-.02in]{$\beta_0$}{$\varphi$} & $1/9$ & $1/3$ & $1$ &   & $1/9$ & $1/3$ & $1$ &   & $1/9$ & $1/3$ & $1$ \\
\hline
\multirow{2}{*}{$3$} & 0 & \cellcolor[HTML]{B1DC3D}{\textcolor{white}{\textbf{-7.9}}} & \cellcolor[HTML]{00A890}{\textcolor{white}{\textbf{-5.0}}} & \cellcolor[HTML]{007094}{\textcolor{white}{\textbf{-3.0}}} &  & \cellcolor[HTML]{194F86}{\textcolor{white}{\textbf{-1.9}}} & \cellcolor[HTML]{44276D}{\textcolor{white}{\textbf{-0.8}}} & \cellcolor[HTML]{491763}{\textcolor{white}{\textbf{-0.4}}} &  & \cellcolor[HTML]{4A095D}{\textcolor{white}{\textbf{-0.2}}} & \cellcolor[HTML]{4B0059}{\textcolor{white}{\textbf{-0.1}}} & \cellcolor[HTML]{4B0058}{\textcolor{white}{\textbf{-0.1}}}\\
 & -4 & \cellcolor[HTML]{D1E02E}{\textcolor{white}{\textbf{-8.4}}} & \cellcolor[HTML]{00B289}{\textcolor{white}{\textbf{-5.4}}} & \cellcolor[HTML]{007595}{\textcolor{white}{\textbf{-3.1}}} &  & \cellcolor[HTML]{264A83}{\textcolor{white}{\textbf{-1.8}}} & \cellcolor[HTML]{44296E}{\textcolor{white}{\textbf{-0.9}}} & \cellcolor[HTML]{481964}{\textcolor{white}{\textbf{-0.5}}} &  & \cellcolor[HTML]{481964}{\textcolor{white}{\textbf{-0.5}}} & \cellcolor[HTML]{4B045B}{\textcolor{white}{\textbf{-0.1}}} & \cellcolor[HTML]{4B0057}{\textcolor{white}{\textbf{0.0}}}\\
\end{tabular}}
\vspace{-.3cm}
\caption{\label{tab:PvSR_CRPS_Admin2_prevalence}Mean percent change in CRPS of the empirical aggregation model relative to the smooth latent aggregation model, where the response is Admin2 prevalence. Yellow-green values are better, while indigo values are worse. Results most representative of the application in Section \ref{sec:application} are outlined in red.}
\end{table}

\begin{table}[h]
\centering
\resizebox{\linewidth}{!}{
\begin{tabular}[t]{c@{\phantom{\tiny a}}c|ccc@{\phantom{a}}p{0em}ccc@{\phantom{a}}p{0em}ccc}
  & $r_{\tiny \mbox{pop}}$ & \multicolumn{3}{c}{$1/5$} &   & \multicolumn{3}{c}{$1$} &   & \multicolumn{3}{c}{$5$} \\[-0.5ex]
$r_{\tiny \mbox{samp}}$ & \diagbox[width=.3in, height=.3in, innerleftsep=.04in, innerrightsep=-.02in]{$\beta_0$}{$\varphi$} & $1/9$ & $1/3$ & $1$ &   & $1/9$ & $1/3$ & $1$ &   & $1/9$ & $1/3$ & $1$ \\
\hline
\multirow{2}{*}{$3$} & 0 & \cellcolor[HTML]{D1E02E}{\textcolor{white}{\textbf{-49.9}}} & \cellcolor[HTML]{00C377}{\textcolor{white}{\textbf{-37.3}}} & \cellcolor[HTML]{009397}{\textcolor{white}{\textbf{-24.9}}} &  & \cellcolor[HTML]{006B93}{\textcolor{white}{\textbf{-16.8}}} & \cellcolor[HTML]{353E7C}{\textcolor{white}{\textbf{-8.8}}} & \cellcolor[HTML]{44296E}{\textcolor{white}{\textbf{-5.0}}} &  & \cellcolor[HTML]{491261}{\textcolor{white}{\textbf{-2.1}}} & \cellcolor[HTML]{4A065C}{\textcolor{white}{\textbf{-1.2}}} & \cellcolor[HTML]{4B025A}{\textcolor{white}{\textbf{-0.8}}}\\
 & -4 & \cellcolor[HTML]{D1E02E}{\textcolor{white}{\textbf{-50}}} & \cellcolor[HTML]{00C475}{\textcolor{white}{\textbf{-37.7}}} & \cellcolor[HTML]{009397}{\textcolor{white}{\textbf{-24.9}}} &  & \cellcolor[HTML]{006490}{\textcolor{white}{\textbf{-15.4}}} & \cellcolor[HTML]{343F7D}{\textcolor{white}{\textbf{-8.9}}} & \cellcolor[HTML]{45266C}{\textcolor{white}{\textbf{-4.8}}} &  & \cellcolor[HTML]{46226A}{\textcolor{white}{\textbf{-4.2}}} & \cellcolor[HTML]{4A065C}{\textcolor{white}{\textbf{-1.2}}} & \cellcolor[HTML]{4B0057}{\textcolor{white}{\textbf{-0.1}}}\\
\end{tabular}}
\vspace{-.3cm}
\caption{\label{tab:PvSR_IntervalScore95_Admin2_prevalence}Mean percent change in 95\% interval score of the empirical aggregation model relative to the smooth latent aggregation model, where the response is Admin2 prevalence. Yellow-green values are better, while indigo values are worse. Results most representative of the application in Section \ref{sec:application} are outlined in red.}
\end{table}

\FloatBarrier
\clearpage
\subsection{Admin2 burden}
\FloatBarrier

\begin{table}[h]
\centering
\resizebox{\linewidth}{!}{
\begin{tabular}[t]{c@{\phantom{\tiny a}}c|ccc@{\phantom{a}}p{0em}ccc@{\phantom{a}}p{0em}ccc}
  & $r_{\tiny \mbox{pop}}$ & \multicolumn{3}{c}{$1/5$} &   & \multicolumn{3}{c}{$1$} &   & \multicolumn{3}{c}{$5$} \\[-0.5ex]
$r_{\tiny \mbox{samp}}$ & \diagbox[width=.3in, height=.3in, innerleftsep=.04in, innerrightsep=-.02in]{$\beta_0$}{$\varphi$} & $1/9$ & $1/3$ & $1$ &   & $1/9$ & $1/3$ & $1$ &   & $1/9$ & $1/3$ & $1$ \\
\hline
\multirow{2}{*}{$3$} & 0 & \cellcolor[HTML]{D1E02E}{\textcolor{white}{\textbf{-18.3}}} & \cellcolor[HTML]{7CD358}{\textcolor{white}{\textbf{-15.7}}} & \cellcolor[HTML]{31C86F}{\textcolor{white}{\textbf{-14.2}}} &  & \cellcolor[HTML]{009597}{\textcolor{white}{\textbf{-9.3}}} & \cellcolor[HTML]{007094}{\textcolor{white}{\textbf{-6.5}}} & \cellcolor[HTML]{00578A}{\textcolor{white}{\textbf{-4.7}}} &  & \cellcolor[HTML]{412E71}{\textcolor{white}{\textbf{-2.2}}} & \cellcolor[HTML]{491662}{\textcolor{white}{\textbf{-1.0}}} & \cellcolor[HTML]{491261}{\textcolor{white}{\textbf{-0.8}}}\\
 & -4 & \cellcolor[HTML]{009796}{\textcolor{white}{\textbf{-9.4}}} & \cellcolor[HTML]{007395}{\textcolor{white}{\textbf{-6.7}}} & \cellcolor[HTML]{194F86}{\textcolor{white}{\textbf{-4.3}}} &  & \cellcolor[HTML]{422C70}{\textcolor{white}{\textbf{-2.1}}} & \cellcolor[HTML]{491662}{\textcolor{white}{\textbf{-0.9}}} & \cellcolor[HTML]{4A0C5E}{\textcolor{white}{\textbf{-0.6}}} &  & \cellcolor[HTML]{4A065C}{\textcolor{white}{\textbf{-0.4}}} & \cellcolor[HTML]{4B0057}{\textcolor{white}{\textbf{-0.1}}} & \cellcolor[HTML]{4B0057}{\textcolor{white}{\textbf{0.0}}}\\
\end{tabular}}
\vspace{-.3cm}
\caption{\label{tab:PvSR_CRPS_Admin2_burden}Mean percent change in CRPS of the empirical aggregation model relative to the smooth latent aggregation model, where the response is Admin2 burden. Yellow-green values are better, while indigo values are worse. Results most representative of the application in Section \ref{sec:application} are outlined in red.}
\end{table}

\begin{table}[h]
\centering
\resizebox{\linewidth}{!}{
\begin{tabular}[t]{c@{\phantom{\tiny a}}c|ccc@{\phantom{a}}p{0em}ccc@{\phantom{a}}p{0em}ccc}
  & $r_{\tiny \mbox{pop}}$ & \multicolumn{3}{c}{$1/5$} &   & \multicolumn{3}{c}{$1$} &   & \multicolumn{3}{c}{$5$} \\[-0.5ex]
$r_{\tiny \mbox{samp}}$ & \diagbox[width=.3in, height=.3in, innerleftsep=.04in, innerrightsep=-.02in]{$\beta_0$}{$\varphi$} & $1/9$ & $1/3$ & $1$ &   & $1/9$ & $1/3$ & $1$ &   & $1/9$ & $1/3$ & $1$ \\
\hline
\multirow{2}{*}{$3$} & 0 & \cellcolor[HTML]{D1E02E}{\textcolor{white}{\textbf{-74.2}}} & \cellcolor[HTML]{B3DC3C}{\textcolor{white}{\textbf{-70.2}}} & \cellcolor[HTML]{9BD848}{\textcolor{white}{\textbf{-67.5}}} &  & \cellcolor[HTML]{00C279}{\textcolor{white}{\textbf{-54.7}}} & \cellcolor[HTML]{00A790}{\textcolor{white}{\textbf{-43.8}}} & \cellcolor[HTML]{008D98}{\textcolor{white}{\textbf{-35.2}}} &  & \cellcolor[HTML]{055489}{\textcolor{white}{\textbf{-18.4}}} & \cellcolor[HTML]{412E71}{\textcolor{white}{\textbf{-8.7}}} & \cellcolor[HTML]{432A6F}{\textcolor{white}{\textbf{-7.8}}}\\
 & -4 & \cellcolor[HTML]{00BE7D}{\textcolor{white}{\textbf{-53.2}}} & \cellcolor[HTML]{00A592}{\textcolor{white}{\textbf{-43.1}}} & \cellcolor[HTML]{007E97}{\textcolor{white}{\textbf{-30.4}}} &  & \cellcolor[HTML]{234B84}{\textcolor{white}{\textbf{-16.0}}} & \cellcolor[HTML]{412E71}{\textcolor{white}{\textbf{-8.7}}} & \cellcolor[HTML]{472069}{\textcolor{white}{\textbf{-5.7}}} &  & \cellcolor[HTML]{491763}{\textcolor{white}{\textbf{-3.9}}} & \cellcolor[HTML]{4B025A}{\textcolor{white}{\textbf{-1.0}}} & \cellcolor[HTML]{4B0057}{\textcolor{white}{\textbf{-0.2}}}\\
\end{tabular}}
\vspace{-.3cm}
\caption{\label{tab:PvSR_IntervalScore95_Admin2_burden}Mean percent change in 95\% interval score of the empirical aggregation model relative to the smooth latent aggregation model, where the response is Admin2 burden. Yellow-green values are better, while indigo values are worse. Results most representative of the application in Section \ref{sec:application} are outlined in red.}
\end{table}

\FloatBarrier
\clearpage
\subsection{Admin2 relative prevalence}
\FloatBarrier

\begin{table}[h]
\centering
\resizebox{\linewidth}{!}{
\begin{tabular}[t]{c@{\phantom{\tiny a}}c|ccc@{\phantom{a}}p{0em}ccc@{\phantom{a}}p{0em}ccc}
  & $r_{\tiny \mbox{pop}}$ & \multicolumn{3}{c}{$1/5$} &   & \multicolumn{3}{c}{$1$} &   & \multicolumn{3}{c}{$5$} \\[-0.5ex]
$r_{\tiny \mbox{samp}}$ & \diagbox[width=.3in, height=.3in, innerleftsep=.04in, innerrightsep=-.02in]{$\beta_0$}{$\varphi$} & $1/9$ & $1/3$ & $1$ &   & $1/9$ & $1/3$ & $1$ &   & $1/9$ & $1/3$ & $1$ \\
\hline
\multirow{2}{*}{$3$} & 0 & \cellcolor[HTML]{50CC68}{\textcolor{white}{\textbf{-22.4}}} & \cellcolor[HTML]{00C574}{\textcolor{white}{\textbf{-21.2}}} & \cellcolor[HTML]{00B884}{\textcolor{white}{\textbf{-19.3}}} &  & \cellcolor[HTML]{00A591}{\textcolor{white}{\textbf{-17.0}}} & \cellcolor[HTML]{009C95}{\textcolor{white}{\textbf{-15.9}}} & \cellcolor[HTML]{008998}{\textcolor{white}{\textbf{-13.8}}} &  & \cellcolor[HTML]{005E8E}{\textcolor{white}{\textbf{-9.6}}} & \cellcolor[HTML]{115187}{\textcolor{white}{\textbf{-8.4}}} & \cellcolor[HTML]{343F7D}{\textcolor{white}{\textbf{-6.9}}}\\
 & -4 & \cellcolor[HTML]{CFE02F}{\textcolor{white}{\textbf{-27.1}}} & \cellcolor[HTML]{CBDF31}{\textcolor{white}{\textbf{-26.9}}} & \cellcolor[HTML]{D1E02E}{\textcolor{white}{\textbf{-27.2}}} &  & \cellcolor[HTML]{00BC80}{\textcolor{white}{\textbf{-19.9}}} & \cellcolor[HTML]{00B983}{\textcolor{white}{\textbf{-19.5}}} & \cellcolor[HTML]{00B785}{\textcolor{white}{\textbf{-19.2}}} &  & \cellcolor[HTML]{006D93}{\textcolor{white}{\textbf{-11}}} & \cellcolor[HTML]{006490}{\textcolor{white}{\textbf{-10.2}}} & \cellcolor[HTML]{006490}{\textcolor{white}{\textbf{-10.2}}}\\
\end{tabular}}
\vspace{-.3cm}
\caption{\label{tab:PvSR_CRPS_Admin2_relativeprevalence}Mean percent change in CRPS of the empirical aggregation model relative to the smooth latent aggregation model, where the response is Admin2 relative prevalence. Yellow-green values are better, while indigo values are worse. Results most representative of the application in Section \ref{sec:application} are outlined in red.}
\end{table}

\begin{table}[h]
\centering
\resizebox{\linewidth}{!}{
\begin{tabular}[t]{c@{\phantom{\tiny a}}c|ccc@{\phantom{a}}p{0em}ccc@{\phantom{a}}p{0em}ccc}
  & $r_{\tiny \mbox{pop}}$ & \multicolumn{3}{c}{$1/5$} &   & \multicolumn{3}{c}{$1$} &   & \multicolumn{3}{c}{$5$} \\[-0.5ex]
$r_{\tiny \mbox{samp}}$ & \diagbox[width=.3in, height=.3in, innerleftsep=.04in, innerrightsep=-.02in]{$\beta_0$}{$\varphi$} & $1/9$ & $1/3$ & $1$ &   & $1/9$ & $1/3$ & $1$ &   & $1/9$ & $1/3$ & $1$ \\
\hline
\multirow{2}{*}{$3$} & 0 & \cellcolor[HTML]{CFE02F}{\textcolor{white}{\textbf{-78.9}}} & \cellcolor[HTML]{BFDE36}{\textcolor{white}{\textbf{-77.3}}} & \cellcolor[HTML]{A8DA42}{\textcolor{white}{\textbf{-74.8}}} &  & \cellcolor[HTML]{89D551}{\textcolor{white}{\textbf{-72.0}}} & \cellcolor[HTML]{73D25B}{\textcolor{white}{\textbf{-70.0}}} & \cellcolor[HTML]{00C475}{\textcolor{white}{\textbf{-64.3}}} &  & \cellcolor[HTML]{00AF8C}{\textcolor{white}{\textbf{-56.7}}} & \cellcolor[HTML]{009E94}{\textcolor{white}{\textbf{-51.8}}} & \cellcolor[HTML]{007B97}{\textcolor{white}{\textbf{-42.5}}}\\
 & -4 & \cellcolor[HTML]{D1E02E}{\textcolor{white}{\textbf{-79.2}}} & \cellcolor[HTML]{CFE02F}{\textcolor{white}{\textbf{-78.8}}} & \cellcolor[HTML]{BFDE36}{\textcolor{white}{\textbf{-77.3}}} &  & \cellcolor[HTML]{9BD848}{\textcolor{white}{\textbf{-73.6}}} & \cellcolor[HTML]{91D74E}{\textcolor{white}{\textbf{-72.7}}} & \cellcolor[HTML]{89D551}{\textcolor{white}{\textbf{-71.8}}} &  & \cellcolor[HTML]{00AF8C}{\textcolor{white}{\textbf{-56.7}}} & \cellcolor[HTML]{00A98F}{\textcolor{white}{\textbf{-55.2}}} & \cellcolor[HTML]{00A592}{\textcolor{white}{\textbf{-53.8}}}\\
\end{tabular}}
\vspace{-.3cm}
\caption{\label{tab:PvSR_IntervalScore95_Admin2_relativeprevalence}Mean percent change in 95\% interval score of the empirical aggregation model relative to the smooth latent aggregation model, where the response is Admin2 relative prevalence. Yellow-green values are better, while indigo values are worse. Results most representative of the application in Section \ref{sec:application} are outlined in red.}
\end{table}

\FloatBarrier
\clearpage
\subsection{Admin1 prevalence}
\FloatBarrier

\begin{table}[h]
\centering
\resizebox{.94\linewidth}{!}{
\begin{tabular}[t]{c@{\phantom{\tiny a}}c|ccc@{\phantom{a}}p{0em}ccc@{\phantom{a}}p{0em}ccc}
  & $r_{\tiny \mbox{pop}}$ & \multicolumn{3}{c}{$1/5$} &   & \multicolumn{3}{c}{$1$} &   & \multicolumn{3}{c}{$5$} \\[-0.5ex]
$r_{\tiny \mbox{samp}}$ & \diagbox[width=.3in, height=.3in, innerleftsep=.04in, innerrightsep=-.02in]{$\beta_0$}{$\varphi$} & $1/9$ & $1/3$ & $1$ &   & $1/9$ & $1/3$ & $1$ &   & $1/9$ & $1/3$ & $1$ \\
\hline
\multirow{2}{*}{$3$} & 0 & \cellcolor[HTML]{214C84}{\textcolor{white}{\textbf{0.2}}} & \cellcolor[HTML]{4B0055}{\textcolor{white}{\textbf{0.7}}} & \cellcolor[HTML]{432A6F}{\textcolor{white}{\textbf{0.4}}} &  & \cellcolor[HTML]{007395}{\textcolor{white}{\textbf{-0.1}}} & \cellcolor[HTML]{006892}{\textcolor{white}{\textbf{0.0}}} & \cellcolor[HTML]{007495}{\textcolor{white}{\textbf{-0.1}}} &  & \cellcolor[HTML]{006791}{\textcolor{white}{\textbf{0.0}}} & \cellcolor[HTML]{006691}{\textcolor{white}{\textbf{0.0}}} & \cellcolor[HTML]{006691}{\textcolor{white}{\textbf{0.0}}}\\
 & -4 & \cellcolor[HTML]{D1E02E}{\textcolor{white}{\textbf{-1.5}}} & \cellcolor[HTML]{00A591}{\textcolor{white}{\textbf{-0.6}}} & \cellcolor[HTML]{0C5288}{\textcolor{white}{\textbf{0.1}}} &  & \cellcolor[HTML]{007896}{\textcolor{white}{\textbf{-0.2}}} & \cellcolor[HTML]{006C93}{\textcolor{white}{\textbf{-0.1}}} & \cellcolor[HTML]{006590}{\textcolor{white}{\textbf{0.0}}} &  & \cellcolor[HTML]{007194}{\textcolor{white}{\textbf{-0.1}}} & \cellcolor[HTML]{006791}{\textcolor{white}{\textbf{0.0}}} & \cellcolor[HTML]{006390}{\textcolor{white}{\textbf{0.0}}}\\
\end{tabular}}
\vspace{-.3cm}
\caption{\label{tab:PvSR_CRPS_Admin1_prevalence}Mean percent change in CRPS of the empirical aggregation model relative to the smooth latent aggregation model, where the response is Admin1 prevalence. Yellow-green values are better, while indigo values are worse. Results most representative of the application in Section \ref{sec:application} are outlined in red.}
\end{table}

\begin{table}[h]
\centering
\resizebox{.94\linewidth}{!}{
\begin{tabular}[t]{c@{\phantom{\tiny a}}c|ccc@{\phantom{a}}p{0em}ccc@{\phantom{a}}p{0em}ccc}
  & $r_{\tiny \mbox{pop}}$ & \multicolumn{3}{c}{$1/5$} &   & \multicolumn{3}{c}{$1$} &   & \multicolumn{3}{c}{$5$} \\[-0.5ex]
$r_{\tiny \mbox{samp}}$ & \diagbox[width=.3in, height=.3in, innerleftsep=.04in, innerrightsep=-.02in]{$\beta_0$}{$\varphi$} & $1/9$ & $1/3$ & $1$ &   & $1/9$ & $1/3$ & $1$ &   & $1/9$ & $1/3$ & $1$ \\
\hline
\multirow{2}{*}{$3$} & 0 & \cellcolor[HTML]{2B4681}{\textcolor{white}{\textbf{-0.1}}} & \cellcolor[HTML]{4B0055}{\textcolor{white}{\textbf{3.4}}} & \cellcolor[HTML]{481B65}{\textcolor{white}{\textbf{2.2}}} &  & \cellcolor[HTML]{0C5288}{\textcolor{white}{\textbf{-0.8}}} & \cellcolor[HTML]{214C84}{\textcolor{white}{\textbf{-0.4}}} & \cellcolor[HTML]{006490}{\textcolor{white}{\textbf{-1.9}}} &  & \cellcolor[HTML]{264A83}{\textcolor{white}{\textbf{-0.3}}} & \cellcolor[HTML]{2B4681}{\textcolor{white}{\textbf{-0.1}}} & \cellcolor[HTML]{274883}{\textcolor{white}{\textbf{-0.2}}}\\
 & -4 & \cellcolor[HTML]{D1E02E}{\textcolor{white}{\textbf{-13.5}}} & \cellcolor[HTML]{009197}{\textcolor{white}{\textbf{-5.0}}} & \cellcolor[HTML]{363D7C}{\textcolor{white}{\textbf{0.4}}} &  & \cellcolor[HTML]{005F8E}{\textcolor{white}{\textbf{-1.6}}} & \cellcolor[HTML]{155087}{\textcolor{white}{\textbf{-0.7}}} & \cellcolor[HTML]{1F4D85}{\textcolor{white}{\textbf{-0.5}}} &  & \cellcolor[HTML]{055489}{\textcolor{white}{\textbf{-0.9}}} & \cellcolor[HTML]{294782}{\textcolor{white}{\textbf{-0.2}}} & \cellcolor[HTML]{294782}{\textcolor{white}{\textbf{-0.2}}}\\
\end{tabular}}
\vspace{-.3cm}
\caption{\label{tab:PvSR_IntervalScore95_Admin1_prevalence}Mean percent change in 95\% interval score of the empirical aggregation model relative to the smooth latent aggregation model, where the response is Admin1 prevalence. Yellow-green values are better, while indigo values are worse. Results most representative of the application in Section \ref{sec:application} are outlined in red.}
\end{table}

\FloatBarrier
\clearpage
\subsection{Admin1 burden}
\FloatBarrier

\begin{table}[h]
\centering
\resizebox{.94\linewidth}{!}{
\begin{tabular}[t]{c@{\phantom{\tiny a}}c|ccc@{\phantom{a}}p{0em}ccc@{\phantom{a}}p{0em}ccc}
  & $r_{\tiny \mbox{pop}}$ & \multicolumn{3}{c}{$1/5$} &   & \multicolumn{3}{c}{$1$} &   & \multicolumn{3}{c}{$5$} \\[-0.5ex]
$r_{\tiny \mbox{samp}}$ & \diagbox[width=.3in, height=.3in, innerleftsep=.04in, innerrightsep=-.02in]{$\beta_0$}{$\varphi$} & $1/9$ & $1/3$ & $1$ &   & $1/9$ & $1/3$ & $1$ &   & $1/9$ & $1/3$ & $1$ \\
\hline
\multirow{2}{*}{$3$} & 0 & \cellcolor[HTML]{294782}{\textcolor{white}{\textbf{0.0}}} & \cellcolor[HTML]{4B0055}{\textcolor{white}{\textbf{0.3}}} & \cellcolor[HTML]{44276D}{\textcolor{white}{\textbf{0.2}}} &  & \cellcolor[HTML]{005F8E}{\textcolor{white}{\textbf{-0.1}}} & \cellcolor[HTML]{214C84}{\textcolor{white}{\textbf{0.0}}} & \cellcolor[HTML]{00578A}{\textcolor{white}{\textbf{-0.1}}} &  & \cellcolor[HTML]{1F4D85}{\textcolor{white}{\textbf{0.0}}} & \cellcolor[HTML]{234B84}{\textcolor{white}{\textbf{0.0}}} & \cellcolor[HTML]{264A83}{\textcolor{white}{\textbf{0.0}}}\\
 & -4 & \cellcolor[HTML]{D1E02E}{\textcolor{white}{\textbf{-1.1}}} & \cellcolor[HTML]{00A293}{\textcolor{white}{\textbf{-0.5}}} & \cellcolor[HTML]{422C70}{\textcolor{white}{\textbf{0.1}}} &  & \cellcolor[HTML]{006390}{\textcolor{white}{\textbf{-0.1}}} & \cellcolor[HTML]{055489}{\textcolor{white}{\textbf{0.0}}} & \cellcolor[HTML]{214C84}{\textcolor{white}{\textbf{0.0}}} &  & \cellcolor[HTML]{00598B}{\textcolor{white}{\textbf{-0.1}}} & \cellcolor[HTML]{1C4E86}{\textcolor{white}{\textbf{0.0}}} & \cellcolor[HTML]{274883}{\textcolor{white}{\textbf{0.0}}}\\
\end{tabular}}
\vspace{-.3cm}
\caption{\label{tab:PvSR_CRPS_Admin1_burden}Mean percent change in CRPS of the empirical aggregation model relative to the smooth latent aggregation model, where the response is Admin1 burden. Yellow-green values are better, while indigo values are worse. Results most representative of the application in Section \ref{sec:application} are outlined in red.}
\end{table}

\begin{table}[h]
\centering
\resizebox{.94\linewidth}{!}{
\begin{tabular}[t]{c@{\phantom{\tiny a}}c|ccc@{\phantom{a}}p{0em}ccc@{\phantom{a}}p{0em}ccc}
  & $r_{\tiny \mbox{pop}}$ & \multicolumn{3}{c}{$1/5$} &   & \multicolumn{3}{c}{$1$} &   & \multicolumn{3}{c}{$5$} \\[-0.5ex]
$r_{\tiny \mbox{samp}}$ & \diagbox[width=.3in, height=.3in, innerleftsep=.04in, innerrightsep=-.02in]{$\beta_0$}{$\varphi$} & $1/9$ & $1/3$ & $1$ &   & $1/9$ & $1/3$ & $1$ &   & $1/9$ & $1/3$ & $1$ \\
\hline
\multirow{2}{*}{$3$} & 0 & \cellcolor[HTML]{005F8E}{\textcolor{white}{\textbf{-1.8}}} & \cellcolor[HTML]{4B0059}{\textcolor{white}{\textbf{1.5}}} & \cellcolor[HTML]{4B0055}{\textcolor{white}{\textbf{1.8}}} &  & \cellcolor[HTML]{2B4681}{\textcolor{white}{\textbf{-0.7}}} & \cellcolor[HTML]{31427E}{\textcolor{white}{\textbf{-0.5}}} & \cellcolor[HTML]{1C4E86}{\textcolor{white}{\textbf{-1.0}}} &  & \cellcolor[HTML]{373C7B}{\textcolor{white}{\textbf{-0.3}}} & \cellcolor[HTML]{3D3576}{\textcolor{white}{\textbf{0.0}}} & \cellcolor[HTML]{3B3778}{\textcolor{white}{\textbf{-0.1}}}\\
 & -4 & \cellcolor[HTML]{D1E02E}{\textcolor{white}{\textbf{-10.1}}} & \cellcolor[HTML]{009A96}{\textcolor{white}{\textbf{-4.6}}} & \cellcolor[HTML]{373C7B}{\textcolor{white}{\textbf{-0.3}}} &  & \cellcolor[HTML]{155087}{\textcolor{white}{\textbf{-1.1}}} & \cellcolor[HTML]{31427E}{\textcolor{white}{\textbf{-0.5}}} & \cellcolor[HTML]{363D7C}{\textcolor{white}{\textbf{-0.4}}} &  & \cellcolor[HTML]{33417E}{\textcolor{white}{\textbf{-0.5}}} & \cellcolor[HTML]{3C3677}{\textcolor{white}{\textbf{-0.1}}} & \cellcolor[HTML]{3A3878}{\textcolor{white}{\textbf{-0.1}}}\\
\end{tabular}}
\vspace{-.3cm}
\caption{\label{tab:PvSR_IntervalScore95_Admin1_burden}Mean percent change in 95\% interval score of the empirical aggregation model relative to the smooth latent aggregation model, where the response is Admin1 burden. Yellow-green values are better, while indigo values are worse. Results most representative of the application in Section \ref{sec:application} are outlined in red.}
\end{table}

\FloatBarrier
\clearpage
\subsection{Admin1 relative prevalence}
\FloatBarrier

\begin{table}[h]
\centering
\resizebox{.999\linewidth}{!}{
\begin{tabular}[t]{c@{\phantom{\tiny a}}c|ccc@{\phantom{a}}p{0em}ccc@{\phantom{a}}p{0em}ccc}
  & $r_{\tiny \mbox{pop}}$ & \multicolumn{3}{c}{$1/5$} &   & \multicolumn{3}{c}{$1$} &   & \multicolumn{3}{c}{$5$} \\[-0.5ex]
$r_{\tiny \mbox{samp}}$ & \diagbox[width=.3in, height=.3in, innerleftsep=.04in, innerrightsep=-.02in]{$\beta_0$}{$\varphi$} & $1/9$ & $1/3$ & $1$ &   & $1/9$ & $1/3$ & $1$ &   & $1/9$ & $1/3$ & $1$ \\
\hline
\multirow{2}{*}{$3$} & 0 & \cellcolor[HTML]{CDE030}{\textcolor{white}{\textbf{-12.8}}} & \cellcolor[HTML]{00BC7F}{\textcolor{white}{\textbf{-9.2}}} & \cellcolor[HTML]{00A392}{\textcolor{white}{\textbf{-7.5}}} &  & \cellcolor[HTML]{005A8C}{\textcolor{white}{\textbf{-3.6}}} & \cellcolor[HTML]{234B84}{\textcolor{white}{\textbf{-2.9}}} & \cellcolor[HTML]{353E7C}{\textcolor{white}{\textbf{-2.3}}} &  & \cellcolor[HTML]{4B045B}{\textcolor{white}{\textbf{-0.3}}} & \cellcolor[HTML]{4A0C5E}{\textcolor{white}{\textbf{-0.5}}} & \cellcolor[HTML]{4A0C5E}{\textcolor{white}{\textbf{-0.5}}}\\
 & -4 & \cellcolor[HTML]{D1E02E}{\textcolor{white}{\textbf{-12.9}}} & \cellcolor[HTML]{A5DA43}{\textcolor{white}{\textbf{-12.0}}} & \cellcolor[HTML]{7ED456}{\textcolor{white}{\textbf{-11.2}}} &  & \cellcolor[HTML]{007094}{\textcolor{white}{\textbf{-4.6}}} & \cellcolor[HTML]{005C8D}{\textcolor{white}{\textbf{-3.7}}} & \cellcolor[HTML]{005589}{\textcolor{white}{\textbf{-3.3}}} &  & \cellcolor[HTML]{4B0055}{\textcolor{white}{\textbf{0.0}}} & \cellcolor[HTML]{4A095D}{\textcolor{white}{\textbf{-0.4}}} & \cellcolor[HTML]{481964}{\textcolor{white}{\textbf{-0.8}}}\\
\end{tabular}}
\vspace{-.3cm}
\caption{\label{tab:PvSR_CRPS_Admin1_relativeprevalence}Mean percent change in CRPS of the empirical aggregation model relative to the smooth latent aggregation model, where the response is Admin1 relative prevalence. Yellow-green values are better, while indigo values are worse. Results most representative of the application in Section \ref{sec:application} are outlined in red.}
\end{table}

\begin{table}[h]
\centering
\resizebox{.999\linewidth}{!}{
\begin{tabular}[t]{c@{\phantom{\tiny a}}c|ccc@{\phantom{a}}p{0em}ccc@{\phantom{a}}p{0em}ccc}
  & $r_{\tiny \mbox{pop}}$ & \multicolumn{3}{c}{$1/5$} &   & \multicolumn{3}{c}{$1$} &   & \multicolumn{3}{c}{$5$} \\[-0.5ex]
$r_{\tiny \mbox{samp}}$ & \diagbox[width=.3in, height=.3in, innerleftsep=.04in, innerrightsep=-.02in]{$\beta_0$}{$\varphi$} & $1/9$ & $1/3$ & $1$ &   & $1/9$ & $1/3$ & $1$ &   & $1/9$ & $1/3$ & $1$ \\
\hline
\multirow{2}{*}{$3$} & 0 & \cellcolor[HTML]{D1E02E}{\textcolor{white}{\textbf{-64.1}}} & \cellcolor[HTML]{6DD15E}{\textcolor{white}{\textbf{-54.1}}} & \cellcolor[HTML]{00C07A}{\textcolor{white}{\textbf{-46.7}}} &  & \cellcolor[HTML]{008B98}{\textcolor{white}{\textbf{-29.7}}} & \cellcolor[HTML]{007295}{\textcolor{white}{\textbf{-23.2}}} & \cellcolor[HTML]{005F8E}{\textcolor{white}{\textbf{-18.6}}} &  & \cellcolor[HTML]{471F68}{\textcolor{white}{\textbf{-4.6}}} & \cellcolor[HTML]{471E67}{\textcolor{white}{\textbf{-4.3}}} & \cellcolor[HTML]{471E67}{\textcolor{white}{\textbf{-4.3}}}\\
 & -4 & \cellcolor[HTML]{CBDF31}{\textcolor{white}{\textbf{-63.5}}} & \cellcolor[HTML]{BBDD38}{\textcolor{white}{\textbf{-61.5}}} & \cellcolor[HTML]{AFDB3E}{\textcolor{white}{\textbf{-60.4}}} &  & \cellcolor[HTML]{009896}{\textcolor{white}{\textbf{-33.3}}} & \cellcolor[HTML]{008A98}{\textcolor{white}{\textbf{-29.5}}} & \cellcolor[HTML]{008297}{\textcolor{white}{\textbf{-27.3}}} &  & \cellcolor[HTML]{4A1060}{\textcolor{white}{\textbf{-2.3}}} & \cellcolor[HTML]{491662}{\textcolor{white}{\textbf{-3.1}}} & \cellcolor[HTML]{45266C}{\textcolor{white}{\textbf{-5.8}}}\\
\end{tabular}}
\vspace{-.3cm}\caption{\label{tab:PvSR_IntervalScore95_Admin1_relativeprevalence}Mean percent change in 95\% interval score of the empirical aggregation model relative to the smooth latent aggregation model, where the response is Admin1 relative prevalence. Yellow-green values are better, while indigo values are worse. Results most representative of the application in Section \ref{sec:application} are outlined in red.}
\end{table}

 \FloatBarrier
 \clearpage
\section{Additional NMR application results}
\label{sec:supplementAdditionalApplication}

\rotatebox{90}{\begin{minipage}{0.9\textheight}
\includegraphics[width=\textwidth]{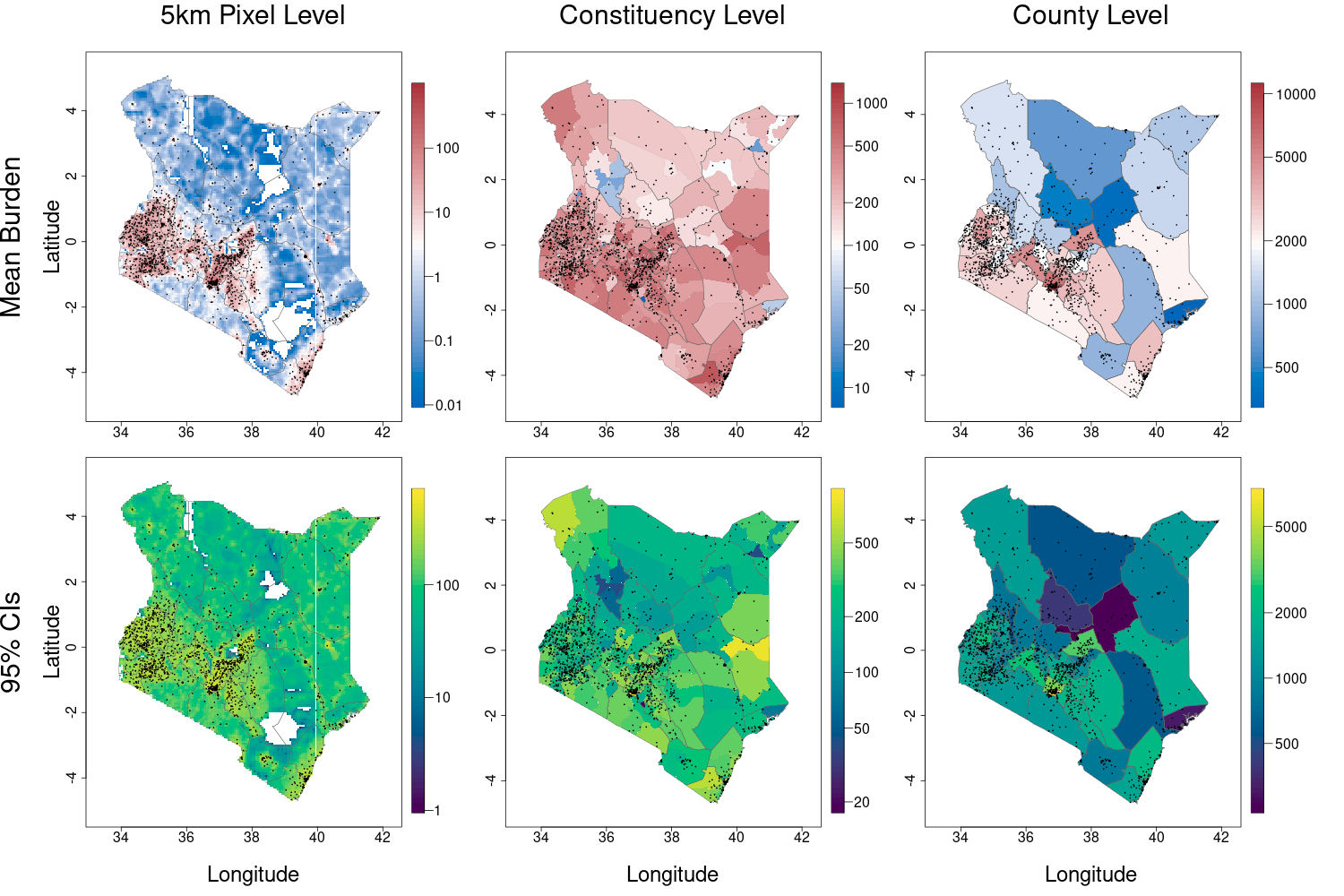}
\captionof{figure}{Central predictions (top row) and 95\% credible interval (CI) widths (bottom row) of neonatal mortality burden in Kenya in 2010--2014. Observation locations are plotted as black dots, provinces as thick black lines, and counties as thin grey lines.}
\label{fig:nmrBurden}
\end{minipage}}

\begin{figure}
\centering
\includegraphics[width=5in]{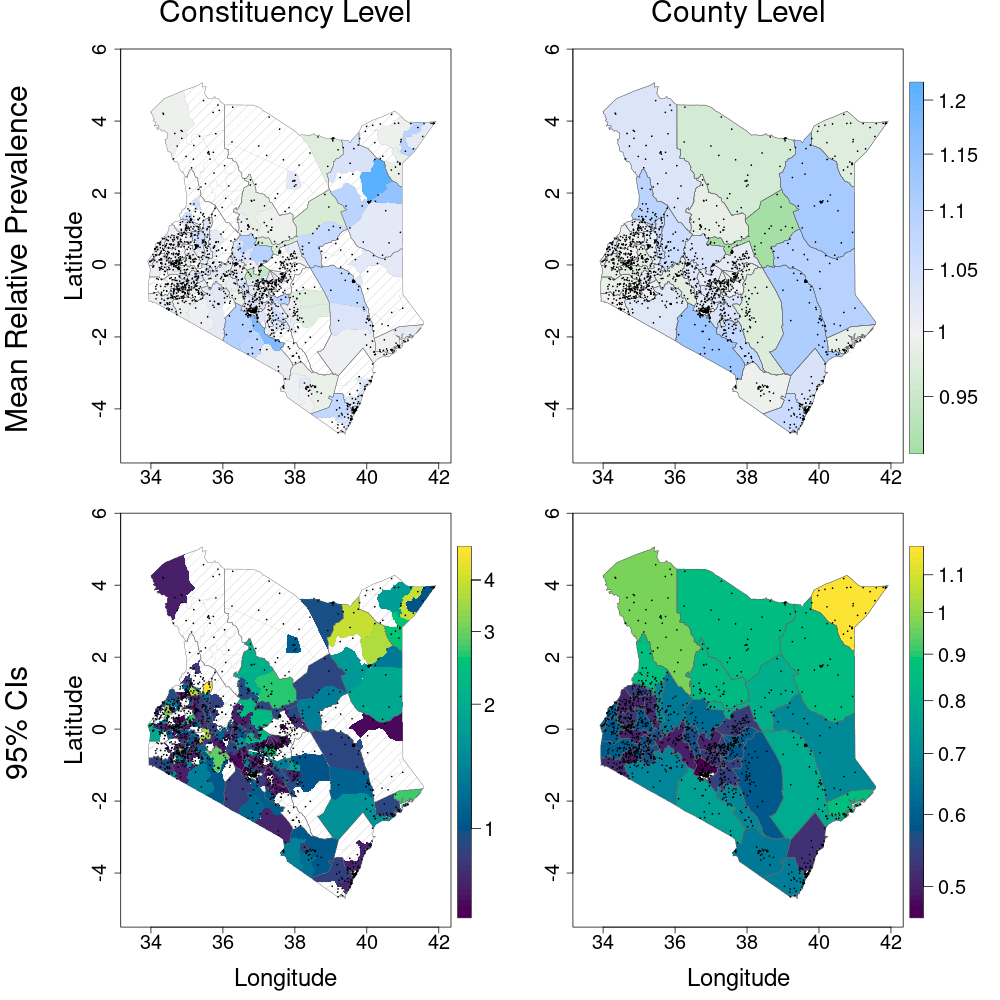}
\caption{Central predictions (top row) and 95\% credible interval widths (bottom row) of neonatal mortality relative prevalence of urban versus rural populations in Kenya in 2010--2014. Observation locations are plotted as black dots, provinces as thick black lines, and counties as thin grey lines. Crosshatched areas do not have both urban and rural populations and so do not have defined relative prevalence.}
\label{fig:nmrRelativePrevalence}
\end{figure}

\bibliographystyle{apalike}

\bibliography{myBib}

\makeatletter\@input{crossReferences1.tex}\makeatother